%% file: phase.tex
\documentstyle[psfig,12pt]{article}
\input lmacros

\def\angle{\mop{angle}}

\def\comment#1{  {\tt [#1]}}
\begin{document}
\baselineskip=15.5pt
\pagestyle{plain}
\setcounter{page}{1}

\begin{titlepage}

\begin{flushright}
PUPT-1936 \\
hep-th/0006119
\end{flushright}
\vfil

\begin{center}
{\huge Phase structure of non-commutative}
\vskip0.5cm
{\huge scalar field theories}
\end{center}

\vfil
\begin{center}
{\large S. S. Gubser and S. L. Sondhi}
\end{center}

$$\seqalign{\span\TL & \span\TT}{
& Joseph Henry Laboratories, Princeton University, Princeton,
NJ 08544
}$$
\vfil

\begin{center}
{\large Abstract}
\end{center}

\noindent 
 We investigate the phase structure of non-commutative scalar field
theories and find evidence for ordered phases which break translation
invariance.  A self-consistent one-loop analysis indicates that the
transition into these ordered phases is first order.  The phase
structure and the existence of scaling limits provides an alternative
to the structure of counter-terms in determining the renormalizability
of non-commutative field theories.  On the basis of the existence of a
critical point in the closely related planar theory, we argue that
there are renormalizable interacting non-commutative scalar field
theories in dimensions two and above. We exhibit this renormalization
explicitly in the large $N$ limit of a non-commutative $O(N)$
vector model.

\vfil
\begin{flushleft}
June 2000
\end{flushleft}
\end{titlepage}
\newpage
\section{Introduction}
\label{Introduction}

Non-commutative field theories have recently received a great deal of
attention \cite{Filk,
cds,SusskindBigatti,SeibergWitten,ChepelevRoiban,Seiberg,GMS,Harvey},
stemming in part from the fact that they arise as low-energy
descriptions of string backgrounds with anti-symmetric tensor fields
\cite{SeibergWitten}.  Renormalizability of non-commutative theories
remains an open question.\footnote{As this paper was nearing
completion, we received \cite{Rivelles}, which argues that a
four-dimensional non-commutative Wess-Zumino model is renormalizable.
The argument hinges on controlling the infrared singularities which
give rise to the interesting phase structure explored in this paper.
Thus there is little overlap between \cite{Rivelles} and our work.}
Standard approaches to demonstrating perturbative renormalizability
(see for instance \cite{ChepelevRoiban}) encounter difficulties
because of infrared singularities.  These singularities are not
associated with any massless propagating fields in the theory, but
instead arise through loop effects.  They can have dramatic physical
consequences: for instance, in a theory whose classical action is that
of a massive scalar with cubic interactions, the $\phi=0$ vacuum
becomes not just globally unstable (on account of an effective
potential which is unbounded below), but locally unstable, as if the
scalar had become tachyonic \cite{Seiberg}.

The main interest in this paper will be in scalar theories with $\phi^4$
interactions.  For the most part we will restrict attention to
Euclidean signature and to even dimensions.  Following \cite{Seiberg},
we will briefly review in Section~\ref{OneLoopDiagrams} how non-planar
one-loop graphs lead to a singularity in the one particle irreducible
(1PI) two-point function: $\Gamma^{(2)}(p) \to \infty$ as $p \to 0$.
What this amounts to physically is long-range frustration: $\langle
\phi(x) \phi(0) \rangle$ oscillates in sign for large $x$.  A natural
expectation, given such a correlator, is that the usual Ising-type
phase transition, to an ordered phase with $\langle \phi \rangle \neq
0$, will be modified to a transition to a phase where $\langle \phi
\rangle$ varies spatially.  This is indeed what we will find in
Section~\ref{Phase}: more particularly, we will find a
fluctuation-driven first order transition to a stripe phase, where
only one momentum mode of the scalar field
condenses.\footnote{In \cite{Campbell} it was argued that a 
translationally invariant ordered phase with massless Goldstone bosons
is impossible in continuum renormalized perturbation theory.  This can
be regarded as a hint of an exotic ordered phase such as the stripe
phase that we find.}  In
Section~\ref{StripesSquares} we will consider more complicated ordered
phases, where more than one momentum mode condenses.  In the
perturbative regime of Section~\ref{Phase}, it turns out that stripe
phases are favored; however, it appears that as couplings are
increased, the system alternates between preferring the condensation
of one or several momentum modes.

The overall picture we will find for the phase diagram is a first
order line terminating on one end at a Lifshitz point, where first
order behavior merges back into the second order transition of the
Ising model; and terminating on the other end at a critical point
which arises in a planar version of the commutative theory.  We review
in Section~\ref{Psandrenorm} the relationship between phase structure
and renormalizability, and argue via scaling that the existence of the
critical point in the planar theory should imply the renormalizability
of the non-commutative theory.

The main method we use to establish the existence of phase transitions
is a self-consistent Hartree treatment of one-loop graphs.  This same
method is our primary tool in demonstrating the validity of the
scaling arguments that guarantee renormalizability. Thus, these
arguments are airtight only for the large $N$ limit of a
non-commutative $O(N)$ vector model, where the self-consistent
one-loop Hartree treatment becomes exact.\footnote{Strictly speaking,
this is true only in the disordered phase and at critical points
at the boundary of this phase. In the ordered phase the distinction
between the self-consistent $N=1$ problem and the large $N$ limit
could be important, see e.g. \cite{Moore} and especially
\cite{Golubovic}. We expect to return to this question in the current
context in the future.}
More sophisticated methods
are called for to decide the validity of scaling relations in more
general non-commutative field theories.  As explained in
Section~\ref{ONVector}, it is possible to arrange the quartic
couplings in the $O(N)$ theory so that there are no divergences at all
at leading order in $N$, independent of the dimension.  Our
renormalizability arguments apply, however, for arbitrary quartic
couplings in the $O(N)$ theory.

Other authors \cite{Fischler:2000fv,Fischler:2000bp,Arcioni:2000bz}
have discussed finite temperature effects in non-commutative field
theories, using the Matsubara formalism with periodic Euclidean time.
The aim of this paper is rather different: we work with
non-commutative field theories on uncompactified flat space, and
varying ``temperature'' is regarded as equivalent to changing the bare
mass.

We conclude in Section~\ref{Concluding} with a discussion of the
relevance of our results to string theory and to quantum hall systems.

\section{Phase Structure and Renormalizability}
\label{Psandrenorm}

We begin with an extremely brief recapitulation of the Wilsonian connection
between the phase structure of a (cutoff) field theory and that of
taking its continuum limit---which is the problem of renormalizability. As
we are interested in this paper in the renormalization of scalar fields,
we will recall the lore on commuting scalar fields.
  \begin{figure}
   \centerline{\psfig{figure=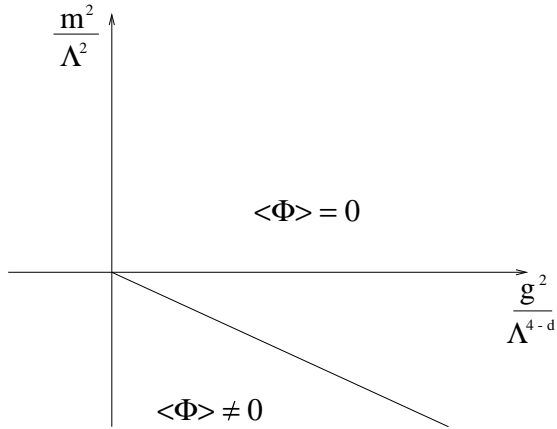,width=2.9in}}
   \caption{The phase diagram for the Ising scalar field
theory,~\eno{CAct}, in $d>1$.}\label{figD}
  \end{figure}

In the Wilsonian approach, we study the Euclidean field theory
governed by the action
  \eqn{CAct}{
   S = \int d^d x \left[ {1 \over 2} (\partial\phi)^2 +
    {1 \over 2} m^2 \phi^2 + {g^2 \over 4} \phi^{4} \right]
  }
 as a statistical mechanics problem---i.e.\ we introduce a momentum
cutoff $\Lambda$ and measure all dimensionful quantities in its
units. This leaves us with a problem with one degree of freedom per
dimensionless volume ($\Lambda^{-d}$ in physical units) and
dimensionless couplings $m^2/\Lambda^2$ and $g^2/\Lambda^{4-d}$,
commonly labelled $r$ and $u$ in the statistical mechanics/condensed
matter literature (see for instance \cite{Chaikin}). We next search
for lines of continuous phase transitions, or critical surfaces, in
the $r,u$ plane; the critical surface in this problem (which is in the
universality class of the Ising model) is sketched in
Figure~\ref{figD}.  On this surface, the correlation length, in
dimensionless units, $\xi$ (the ``lattice'' correlation length,
literally so if the cutoff is implemented via discretization) diverges
which is the {\it sine qua non} of taking a continuum limit. Having
located a critical surface, three continuum limits are possible at
each point on it---a massless limit obtained by sitting exactly on the
critical surface, and two massive limits in which $r(\Lambda)$ and
$u(\Lambda)$ are chosen to approach the critical surface from either
phase as the cutoff is taken to infinity while keeping
$\Lambda/\xi(r,u)=M_R$ fixed and equal to a renormalized mass. That
such limits can be taken, requires {\it scaling} in the statistical
mechanics. Also, the phenomenon of universality will imply that some
domain of a critical surface will exhibit the same long distance
correlations and hence give rise to the {\it same} continuum limit.
  \begin{figure}
   \centerline{\psfig{figure=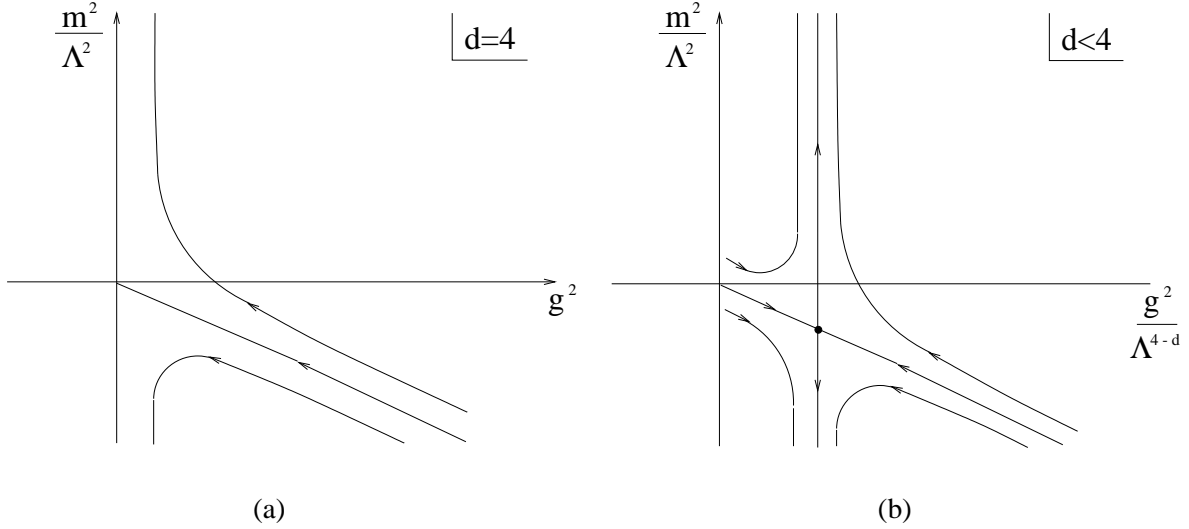,width=6.2in}}
   \caption{Renormalization group flows for the Ising scalar field
theory in (a) $d=4$ and (b) $d < 4$.}\label{figE}
  \end{figure}

In the above description we have used the language of phase
structure. A more powerful account is that of renormalization group
(RG) flows which we have sketched in $d=4$ (Figure~\ref{figE}a) and
$d<4$ (Figure~\ref{figE}b). In the RG description we are interested in
fixed points of infinite correlation length. In $d=4$ we see that all
critical theories flow into the gaussian fixed point $(r,u)=(0,0)$
whence the (strict) continuum limit is trivial, while in $d<4$ the
non-trivial Wilson-Fisher fixed point leads to an interacting
continuum limit. This information is not available from a phase
diagram alone. The fixed point analysis also implies scaling and hence
guarantees a continuum limit.

We belabor this point because in the balance of the paper, we will
deal in phase diagrams and not RG flows. We will find critical points
and will argue that these guarantee the existence of continuum limits
(renormalizability) of non-commuting scalar theories. In a specific
large $N$ theory, we will be able to show this explicitly, and we do
not doubt that the claim is correct. Nevertheless, it does not have
the full generality of an RG analysis and it would be nice to carry
out such an analysis, even perturbatively as done for the commutative
$\phi^4$ theory in $d=4$ by Polchinski \cite{Polchinski84}.

\section{One Loop Action for Non-Commutative Scalars}
\label{OneLoopAction}

\subsection{Generalities}
\label{Generalities}

Our starting point is the non-commutative scalar field theory
specified by the action
  \eqn{NCAct}{
   S = \int d^d x \left[ {1 \over 2} (\partial\phi)^2 + 
    {1 \over 2} m^2 \phi^2 + {g^2 \over 4} \phi^{\star 4} \right]
  }
 where $\phi^{\star 4} = \phi\star\phi\star\phi\star\phi$ and the star
product is defined as usual by
  \eqn{StarProduct}{
   (f \star g)(x) = \left. e^{{i \over 2} \Theta^{\mu\nu} 
    \partial_{x^\mu} \partial_{y^\nu}} f(x) g(y) \right|_{y=x} \,.
  }
 The anti-symmetric matrix $\Theta^{\mu\nu}$, whose relation to the
star product can be most simply expressed through
  \eqn{xComs}{
   [x^\mu,x^\nu]_\star \equiv x^\mu \star x^\nu - x^\nu \star x^\mu
    = i \Theta^{\mu\nu} \,,
  }
 will for most of this paper be assumed to be of the form
  \eqn{ThetaAssume}{
   \Theta^{\mu\nu} = \theta \pmatrix{ 0 & 1 \cr -1 & 0} \otimes
    {\bf 1}_{d/2}
  }
 for even dimensions $d$.  We will comment in
Section~\ref{OddDimensions} on the more complicated cases where
$\Theta^{\mu\nu}$ has unequal eigenvalues or $d$ is odd.

 A useful extension of \StarProduct\ is 
  \eqn{MultiStar}{
   (f_1 \star f_2 \star \ldots \star f_\ell)(x) = 
    \left. e^{{i \over 2} \sum_{i<j} \Theta^{\mu\nu} 
    \partial_{x_i^\mu} \partial_{x_j^\nu}} f_1(x_1) f_2(x_2) 
     \cdots f_\ell(x_\ell) \right|_{x_i=x} \,.
  }
 The result \MultiStar\ is easiest to obtain via Fourier analysis,
using the basic result
  \eqn{ExpStar}{
   e^{i p_1 \cdot x} \star e^{i p_2 \cdot x} = 
    e^{-{i \over 2} p_1 \wedge p_2} e^{i (p_1 + p_2) \cdot x} \,,
  }
 where by definition $p \wedge q = \Theta^{\mu\nu} p_\mu p_\nu$.
(Condensed matter readers should note that this is the lowest Landau
level algebra of density operators \cite{Girvin86}).

Like matrix multiplication, the star product is non-commutative.
However, a product of exponentials $e^{i p_k \cdot x}$, can be
reordered cyclically if the $p_k$ sum to zero.  As a result it is
necessary to specify the cyclic order of vertices in writing down
Feynman rules, just as in large $N$ theories.  It is well known
\cite{Filk} that planar amplitudes of the field theory \NCAct\ are
independent of $\Theta^{\mu\nu}$ (and hence are the same as in the
commutative theory where $\Theta^{\mu\nu}=0$), up to an overall phase.
The effects of non-planar diagrams were first studied systematically
in \cite{Seiberg}.

We will also consider two variants of \NCAct, namely complex scalars,
where
  \eqn{ComplexAct}{
   S = \int d^d x \, \left[ \partial\phi \partial\phi^* + 
    m^2 \phi \phi^* + g^2
     \phi\star\phi^*\star\phi\star\phi^* + 
    g'^2 \phi\star\phi\star\phi^*\star\phi^* \right] \,,
  }
 and the non-commutative $O(N)$ vector model,
  \eqn{ONAct}{
   S = \int d^d x \, \left[ {1 \over 2} (\partial\phi_i)^2 + 
    {1 \over 2} m^2 \phi_i^2 + {g^2 \over 4} \phi_i \star \phi_i 
     \star \phi_j \star \phi_j + 
    {g'^2 \over 4} \delta^{ik} \delta^{jl} 
      \phi_i \star \phi_j \star \phi_k \star \phi_l \right]
  }
 The field theory \ComplexAct\ has been studied previously
\cite{ArefOne}.

\subsection{One loop diagrams}
\label{OneLoopDiagrams}

In this section we review the results of \cite{Seiberg} which will be
relevant for our calculations.  The one loop corrections to the
propagator of the scalar field $\phi$ split into planar and non-planar
parts:
  \eqn{PlanarNonPlanar}{\eqalign{
   \Sigma_{\rm planar}(p) &= 2 g^2
    \int {d^d k \over (2\pi)^d} {1 \over k^2 + m^2}  \cr
   \Sigma_{\rm non-planar}(p) &= g^2
    \int {d^d k \over (2\pi)^d} {e^{i p \wedge k}
     \over k^2 + m^2} \,.
  }}
 Using the Schwinger parametrization 
  \eqn{Schwinger}{
   {1 \over k^2 + m^2} = \int_0^\infty d\alpha\, 
    e^{-\alpha (k^2+m^2)} \,,
  }
 one easily extracts 
  \eqn{NonPlanarG}{\eqalign{
   I_2(p) &= \int^\Lambda {d^d k \over (2\pi)^d} 
     {e^{i p \wedge k} \over k^2+m^2}  \cr
     &= {1 \over (4\pi)^{d/2}} 
      \int_0^\infty {d\alpha \over \alpha^{d/2}} 
      e^{-\alpha m^2 - {p \circ p \over 4\alpha} - 
       {1 \over \Lambda^2 \alpha}}  \cr
     &= (2\pi)^{-d/2} m^{{d-2 \over 2}}
       \left( p \circ p + {4 \over \Lambda^2} \right)^{{2-d \over 4}}
        K_{{d-2 \over 2}}\left( m \sqrt{p \circ p + {4 \over
         \Lambda^2}} \right) \,,
  }}
 where, following \cite{Seiberg}, we have introduced an ultraviolet
regulator $\Lambda$ and defined a symmetric product
  \eqn{CircProd}{
   p \circ q = -p_\mu \Theta^{\mu\nu} \Theta_\nu{}^\lambda q_\lambda
    \,.
  }
 $K_\nu(x)$ denotes a modified Bessel function.

The end result is a corrected propagator of the following form:
  \eqn{EndProp}{
   \Gamma^{(2)}(p) = p^2 + m^2 + 2 g^2 I_2(0) + g^2 I_2(p) \,.
  }
 The parallel results for the complex scalar and the $O(N)$ vector
model are
  \eqn{ComplexON}{\eqalign{
   \hbox{complex scalar:} \qquad & \Gamma^{(2)}(p) = 
     p^2 + m^2 + (2g^2 + g'^2) I_2(0) + g'^2 I_2(p)  \cr
   \hbox{$O(N)$ model:} \qquad & \Gamma^{(2)}(p) = 
     p^2 + m^2 + g^2 N I_2(0) + 
      g'^2 N I_2(p) + O(1) \,,
  }}
 where for the $O(N)$ model we have only evaluated the graphs which
contribute at leading order in large $N$.  The salient property of
$\Gamma^{(2)}(p)$ for our subsequent discussion is that it grows for
small $p$.  If we think of absorbing the planar one-loop contribution
to $\Gamma^{(2)}(p)$ into the definition of mass (for instance, define
a renormalized mass through $M^2 = m^2 + 2 g^2 I_2(p^2)$ for the real
scalar theory), and then removing the cutoff while holding $M^2$
fixed, then $\Gamma^{(2)}(p)$ is finite for finite $p$ but goes to
$+\infty$ as $p \to 0$.  This means that the low-momentum modes are
extremely stiff; thus they seem likely never to participate in a phase
transition.  If there is a phase transition, it should involve the
momentum modes where $\Gamma^{(2)}(p)$ is smallest, since these are
the ones most likely to be destabilized as $M^2$ becomes negative.
The details of how this happens turn out to be somewhat intricate, and
the next section is devoted to sorting them out.

\section{Phase Structure for Non-commutative scalars}
\label{Phase}

\subsection{Generalities}

\begin{figure}
   \centerline{\psfig{figure=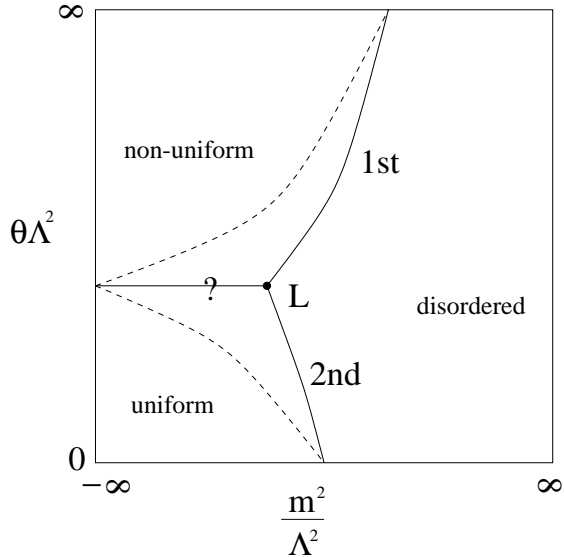,width=2.9in}}
   \caption{The phase diagram for the non-commutative $\phi^4$ theory
at fixed $g^2$, in $d=4$.  The solid lines show transitions for
the case of a single real scalar.  The point marked L is a Lifshitz
point.  The transition between uniform and non-uniform ordered phases
is not accessible, in $d=4$, via the techniques used in this paper.  
Instead, the dashed lines represent the transitions that arise in a 
self-consistent one-loop analysis.}
\label{figG}
\end{figure}

The non-commutative action \NCAct\ is characterized by three
dimensionless parameters: $m^2/\Lambda^2$, $g^2/\Lambda^{4-d}$ and
$\theta \Lambda^2$. Accordingly we need to establish a phase diagram
in a three dimensional space. In what follows we will typically work
at a fixed dimensionless coupling and take two dimensional cuts
through parameter space instead. The resulting phase diagram is
sketched in Figure~\ref{figG}, and the rest of this section is devoted
to the arguments that give rise to it.

\subsection{The Planar Limit: $\theta \Lambda^2=\infty$}

It has been noted previously that the large $\theta$ limit picks out
planar diagrams \cite{Filk}.  In the context of a cutoff field theory,
this statement can be made precise.

Consider the theory at $m^2/\Lambda^2 > 0$ so that all diagrams
are infrared finite in addition to being ultraviolet finite
on account of the cutoff. The perturbative expansion for the
free energy $F= - \log Z$ is the sum of all connected vacuum
diagrams (no external legs). Of these, the planar graphs
involve no phase factors stemming from the non-commutativity
while the non-planar graphs involve at least one. In
the limit $\theta \Lambda^2 \rightarrow \infty$ the latter
vanish on account of the infinitely rapid oscillations of
the integrand. Consequently, the high temperature expansion
of $F$ in the maximally non-commuting scalar theory is
identical to that of the planar theory, defined as the sum
over planar diagrams alone.\footnote{For us, high temperature is
synonymous with large positive $m^2$, while low temperature means
large negative $m^2$.}

This statement can be extended to selected classes of fields.
The perturbative expansion for the free energy in a field is,
\eqn{FreeEn}{\eqalign{
-F[J] &= \log Z[J] \cr
      &= \log Z[0] + \sum_n \int_{p_1} \cdots \int_{p_n}
G_c(p_1,....,p_n) J(p_1) .... J(p_n)
}}
where $G_c(p_1,....,p_n)$, inclusive of the momentum conserving
delta function, is the connected $n$-point Green function in
momentum space. On account of the vanishing of the non-planar
graphs alluded to earlier, 
\eqn{GreenFnEq}{
\lim_{\theta \Lambda^2 \rightarrow \infty} e^{{i\over2}
\sum_{i<j} p_i \wedge p_j} G_c(p_1,....,p_n; \theta \Lambda^2)
= G_c^{\rm planar}(p_1,....,p_n)
}
for each $n$.
For fields that are modulated only in one
direction, the momenta $p_1$ through $p_n$ are parallel whence
the explicit phase factor for planar diagrams vanishes. Hence
in the limit $\theta \Lambda^2 \rightarrow \infty$, $F[J]$
is still given as a sum of planar diagrams alone. Note that
for more complicated field configurations, the equivalence
is no longer true.

Thus far we have made statements in the high temperature phase.  The
planar theory, inheriting the Ising symmetry ($\phi \to -\phi$) of the
full theory, will exhibit a twofold degenerate broken symmetry phase.
This will be reached on traversing a continuous phase transition at a
critical $m_c^2/\Lambda^2 < 0$ which will be mean field in $d>4$, mean
field corrected by logarithms in $d=4$, but likely in a different
universality class from the standard Ising transition in $d<4$ (see
below).  An immediate deduction from the equivalence of the infinite
$\theta \Lambda^2$ and planar theories in a field is that we may
analytically continue their common free energy into the low
temperature phase (bypassing the critical point), thereby establishing
the equivalence of their low temperature thermodynamics as
well. Altogether, we may conclude that the infinite $\theta \Lambda^2$
theory has a critical point at the same $m_c^2/\Lambda^2$ as the
planar theory, which is exactly the same transition. This critical
point will play a central role in achieving a continuum limit for the
noncommutative theory.

Finally we should note the well known result that the planar theory is
also the $N \rightarrow \infty$ limit of a hermitian matrix model with
an appropriately generalized version of \CAct\ as its action.  Via
this route, there are exact results on the phase transition in $d=1$
\cite{GrossMiljkovic}\footnote{It is common to refer to the $c=1$
matrix model as string theory in two dimensions, on account of the
anomaly-induced dynamics for the Liouville field.  From the point of
view of the matrix model, however, $d=1$.} that show that the
transition is different from that of the Ising model, and an
approximate RG treatment that has yielded exponents in $2<d<4$ as well
\cite{Nishigaki,Ferretti}. There does not appear to be a treatment of the
broken symmetry phase in this formulation of the problem---at issue is
how the Ising symmetry of the planar theory is embedded in what
appears {\it prima facie} to be a much larger symmetry group in the
matrix model.

\subsection{First Order Transition at Large $\theta \Lambda^2$}
\label{FirstOrder}

Having established that there is a critical point at $\theta \Lambda^2
=\infty$, we will now show that this terminates a line of fluctuation
driven first order transitions at large $\theta \Lambda^2$. That this
should happen, is a fairly general expectation from the work of
Brazovskii \cite{Brazovskii} when combined with the observation
\cite{Seiberg} that the one loop $\Gamma^{(2)}(p)$ has a minimum at
non-zero $p$ at large $\theta \Lambda^2$ (the latter condition is
implicit in their analysis), and we will review this physics
below. Following Brazovskii, we will employ the self-consistent
Hartree approximation.  This approximation is sensible in $d \geq 4$,
but does not capture the planar theory transition in $d=2$.
Consequently, we will not treat that case in detail, although our
later arguments on renormalizability will apply there as well.

We will present two separate arguments for the first order transition.
The first will involve an asymptotic construction of the solution to
the self-consistent problem---this will follow the original analysis
as closely as possible but with complications that we detail below.
This will also require that we take a double limit in which 
$\theta \Lambda^2$ is taken large but $g^2$ is taken to zero---i.e.
we will be working in the vicinity of the massless Gaussian theory.
The second will be a more indirect argument which will be carried out
at fixed $g^2$ and will consist of showing that the free energy of
two solutions {\it must} cross in a certain region of parameter space.

\subsubsection{Take I}
\label{TakeI}

We turn to the direct construction. To keep the discussion simple, we 
will continue to
assume even Euclidean dimension $d$, and $\Theta^{\mu\nu}$ with
maximal rank and eigenvalues $\pm\theta$.  We will also omit
inessential factors of order unity.

\begin{figure}
   \centerline{\psfig{figure=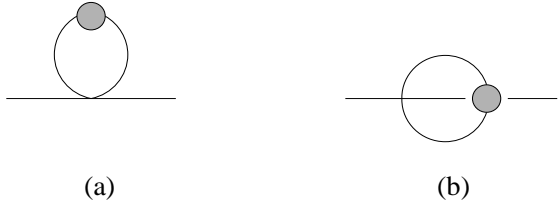,width=2.9in}}
   \caption{1PI graphs with dressed propagators.}\label{figB}
  \end{figure}

A self-consistent treatment of the one-loop correction to the
propagator leads to
  \eqn{SelfGamma}{
   \Gamma^{(2)}(p) = p^2 + m^2 + 
    2 g^2 \int {d^d k \over (2\pi)^d} 
     {1 + {1 \over 2} e^{i k \wedge p} \over \Gamma^{(2)}(k)}
  }
 Having $\Theta^{\mu\nu}$ of maximal rank, with eigenvalues
$\pm\theta$, leads to a considerable simplification: $\Gamma^{(2)}(p)$
is a function only of $p^2$.  This can be shown inductively, order by
order in $g^2$.  Suppose that $\Gamma^{(2)}(p)$ is $SO(d-1)$ to
$\ell$-th order in $g^2$.  The $(\ell+1)$-st order expression for
$\Gamma^{(2)}(p)$ is then invariant on $SO(d-1)$ transformations of
$p_\mu \Theta^{\mu\nu}$, and so can only be a function of $p \circ p =
\theta^2 p^2$.  Equation~\SelfGamma\ can thus be re-expressed as
  \eqn{OneDimGamma}{
   \Gamma^{(2)}(p) = p^2 + M^2 + g^2 \int_0^\infty k^{d-1} dk
     {J_{(d-2)/2}(\theta k p) \over (\theta k p)^{(d-2)/2}} 
     {1 \over \Gamma^{(2)}(k)} \,.
  }
 Here $m^2$ is the bare mass, and 
  \eqn{CorrectedM}{
   M^2 = m^2 + 2 g^2 \int {d^d k \over (2\pi)^d} {1 \over \Gamma^{(2)}(k)}
  }
 is the mass corrected by the planar graph, Figure~\ref{figB}a.  The
non-commutativity does not affect the large $p$ behavior,
$\Gamma^{(2)}(p) \sim p^2$.  Thus the integral in \CorrectedM\ must be
regulated, for instance with an explicit cutoff as in
Section~\ref{OneLoopDiagrams}.  No other ultraviolet divergences will
arise in the following computations, so we may in effect take the
cutoff to infinity and treat $M^2$ as the finite quantity which we
dial to produce a transition.  This leads to an argument for
renormalizability, as we shall explain in Section~\ref{Continuum}.

The last term in \SelfGamma\ has a contribution from large momenta $k$
which goes as $1/p^{d-2}$ as $p \to 0$.  This prevents condensation of
very low-momentum modes of $\phi$.  Thus, if there is a phase
transition at all, {\it the ordered state must break translation
invariance}.

From now on, let us restrict attention to $d=4$.  At the end of this
section we will remark on the extension to other dimensions.
Proceeding naively, we could solve \OneDimGamma\ to the first
non-trivial order in $g$ by replacing $\Gamma^{(2)}(k)$ on the right
hand side by $k^2+M^2$.  Then as $M^2$ is decreased below $-c_1 {g
\over \theta}$, for some (easily computable) constant $c_1$ of order
unity, $\Gamma^{(2)}(p)$ becomes negative near $p = p_c \sim
\sqrt{g/\theta}$, signalling a second order phase transition to an
ordered state where some momentum modes of $\phi$ condense.  Our aim
in the rest of this subsection is to show that this naive result is
altered in a self-consistent analysis to a rather more interesting
conclusion: there is a {\it first} order transition to a stripe
pattern at $M^2 = -c_1 {g \over \theta} - c_2 {g^{7/3} \over \theta}$,
where $c_2$ is another constant of order unity.

First observe that \OneDimGamma\ makes the second order transition
impossible.  If we were to suppose that $\Gamma^{(2)}(p_c) = 0$ for
some $M^2$ and some $p_c$, then the integral in \OneDimGamma\
diverges.\footnote{To be more precise, the integral in \OneDimGamma\
diverges if $\Gamma^{(2)}(p) \sim (p^2-p_c^2)^2$ for $p$ near $p_c$:
that is, $\Gamma^{(2)}(p)$ cannot at the same time remain smooth and
have a zero.  To cement the conclusion that there are no second order
phase transitions, one must check that it is also impossible for
$\Gamma^{(2)}(p)$ to have a cusp form such as $|p^2-p_c^2|^{2\nu}$
with $\nu < 1$.  Such a form can eliminate the divergence in
\OneDimGamma.  But there is still a contradiction:
$d\Gamma^{(2)}(p)/dp$ diverges as $p \to p_c$, but an integral
expression for $d\Gamma^{(2)}(p)/dp$ remains finite.}  The possibility
of a first order transition was first realized by Brazovskii
\cite{Brazovskii}, given a suitable effective Lagrangian, with a
minimum in the inverse propagator at non-zero $p_c$.  In
\cite{Brazovskii}, such an effective Lagrangian was simply assumed,
and a one-loop fluctuation analysis with this starting point was shown
to lead to a first order transition.  This same method was further
developed in \cite{SwiftHTwo}.  In our case, the basic Lagrangian does
{\it not} have an appropriate inverse propagator with a minimum at
nonzero $p_c$---as we have seen, this only arises in the 1PI effective
action after a one-loop analysis.  One cannot feed $\Gamma^{(2)}(p)$
directly into a Brazovskiian analysis: all fluctuations are supposed
to be incorporated into $\Gamma^{(2)}(p)$ already.  However, we shall
see that a self-consistent treatment based on \OneDimGamma\ reproduces
the essential features of \cite{Brazovskii}, and does support the
conclusion that a first order transition takes place.  

Although $\Gamma^{(2)}(p)$ can never vanish, it should be possible to
make its minimum as small as we please by decreasing $M^2$
sufficiently.  We then have the approximate forms
  \eqn{GammaApproximate}{\seqalign{\span\TL & \span\TR \qquad & \span\TT}{
   \Gamma^{(2)}(p) &\approx p^2 & for large $p$  \cr
   \Gamma^{(2)}(p) &\approx \xi_0^2 (p^2-p_c^2)^2 + r & for 
     $p \approx p_c$  \cr
   \Gamma^{(2)}(p) &\approx {g^2 \over (\theta p)^2} &
     for small $p$.
  }}
 Two regions of the integral in \OneDimGamma\ will then make
significant contributions: the large $p$ region and the $p \approx
p_c$ region.  The full integral can be approximated as a sum of the
contributions from these regions.  The parameters $p_c$ and $\xi_0$
can then be determined self-consistently: for $p \approx p_c$, 
  \eqn{SelfXi}{\eqalign{
   \Gamma^{(2)}(p) &\approx p^2 + M^2 + 
     g^2 \left( \int_0^\infty k^3 dk 
      {J_1(\theta k p) / \theta k p \over k^2} + 
       p_c^2 {J_1(\theta p_c^2) \over \theta p_c^2}
       \int_0^\infty {k dk \over r + \xi_0^2 (k^2-p_c^2)^2} \right)  \cr
     &\approx p^2 + M^2 + {g^2 \over (\theta p)^2} +
      {g^2 p_c^2 \over \xi_0 \sqrt{r}}  \cr
   p_c &= {1 \over \xi_0} = \sqrt{g \over \theta} \qquad
   r = 2 p_c^2 + M^2 + 
    {g^3 / \theta \over \xi_0 \sqrt{r}} \,.
  }}
 The last line of \SelfXi\ follows from matching the previous line to
the approximate form $\Gamma^{(2)}(p) \approx \xi_0^2 (p^2-p_c^2)^2$.
We have again suppressed inessential factors of order unity in
\SelfXi, and we have assumed $g \ll 1$, so that $\theta p_c^2 \ll 1$.
This latter inequality will be essential to our further analysis.  It
turns out to be difficult to get the factors of order unity right,
because $\xi_0$ is only on the order of $1/p_c$: the breadth of the
minimum of $\Gamma^{(2)}(p)$, in the approximation scheme we have
used, is comparable to its distance from the origin.

So far we have done all calculations as expansions around the
disordered phase, where $\langle \phi \rangle = 0$.  The putative
ordered phase is a stripe pattern: $\langle \phi \rangle = \phi_0 = A
\cos p_c x$.  (Strictly speaking, we do not expect an exact $\cos p_c
x$ dependence, only a function of $x$ with period $2\pi / p_c$.  But
because the momentum modes near $p_c$ make the dominant contribution,
neglecting higher harmonics should not change the qualitative picture.
One should in principle consider sums of the form $\cos p_c x + \cos
p_c y$, as well as more complicated superpositions; however, as we
shall see in Section~\ref{StripesSquares}, the simple stripe solution
is favored at small coupling).  Writing $\phi = \phi_0 + \eta$ and
expanding in fluctuations of $\eta$, one obtains
  \eqn{LExpanded}{\eqalign{
   {\cal L} &= {1 \over 2} (\partial\phi_0)^2 + 
    {g^2 \over 4} \phi_0^{\star 4} 
    + {1 \over 2} (\partial\eta)^2 + {g^2 \over 4} \eta^{\star 4}  \cr
    &\qquad{} + \partial\phi_0 \partial\eta 
    + g^2 \phi_0^{\star 3} \star \eta 
    + g^2 \phi_0 \star \eta^{\star 3}
    + g^2 \phi_0^{\star 2} \star \eta^{\star 2} +
      {g^2 \over 2} \phi_0 \star \eta \star \phi_0 \star \eta \,.
  }}
 We will treat perturbatively all terms with factors of both $\eta$
and $\phi_0$.  This is justified if $A$ is small, which is not
completely obvious given that the transition is first order.  But for
small $g$, $A$ {\it is} small very close to the phase transition---at
least, this is an assumption which can easily be verified at the end
of the computation.
  \begin{figure}
   \centerline{\psfig{figure=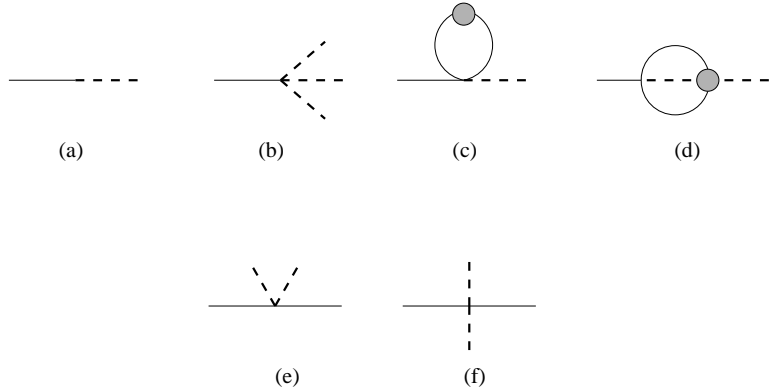,width=4in}}
   \caption{Additional 1PI graphs for the ordered phase.  Dashed
external lines indicate insertions of the background field
$\phi_0$.}\label{figC}
  \end{figure}
 Classically, one would demand that the terms in \LExpanded\ linear in
$\eta$ should vanish: this corresponds to summing the graphs in
Figure~\ref{figC}a and~\ref{figC}b.  At the one-loop level, the graphs
in Figure~\ref{figC}c and~\ref{figC}d also contribute, and the end
result is
  \eqn{hVar}{
   h \equiv {1 \over 2} {d\Phi \over dA} = \Gamma^{(2)}(p_c) A - 
    {g^2 \over 2} A^3 \,,
  }
 where $\Phi$ is the free energy per unit volume.  Adding the graphs
in Figures~\ref{figC}e and~\ref{figC}f to the self-consistent one-loop
graphs that contributed to \SelfGamma, one finds only a slight
modification of the propagator in the disordered
phase:\footnote{Actually, the propagator is not diagonal in momentum
space once the graphs in Figure~\ref{figC}e and~\ref{figC}f are added.
Non-diagonal terms would have to be taken into account if we wanted to
determine the exact form of the ordered phase; however they should not
change the qualitative features of the phase transition.}  To
\SelfGamma\ one must simply add the terms $g^2 A^2 (1 + {1 \over 2}
\cos p \wedge p_c)$.  The first of these terms comes from the graph in
Figure~\ref{figC}e, while the second is from the graph in
Figure~\ref{figC}f.  For momenta $p$ on the order of $p_c$, the
argument of the cosine is small (again assuming $\theta p_c^2$ is
small), so for such momenta we may alter \SelfXi\ to
  \eqn{SelfXiA}{\eqalign{
   \Gamma^{(2)}(p) &\approx p^2 + M^2 + {g^2 \over (\theta p)^2} +
      {g^2 p_c^2 \over \xi_0 \sqrt{r}} + g^2 A^2  \cr
   p_c &= {1 \over \xi_0} = \sqrt{g \over \theta} \qquad
   r = 2 p_c^2 + M^2 + 
    {g^3 / \theta \over \xi_0 \sqrt{r}} + g^2 A^2 \,.
  }}
 
To summarize the situation so far: a self-consistent analysis of the
graphs in Figures~\ref{figB} and~\ref{figC} led to an expression for
the minimum, $r = \Gamma^{(2)}(p_c)$, of the 1PI propagator
$\Gamma^{(2)}(p)$; and an expression for the tadpole for the magnitude
$A$ of the condensate.  These expressions are
  \eqn{ExprA}{
   r = \tau + g^2 A^2 + {\alpha \over \sqrt{r}} \qquad
   {1 \over 2} {\partial\Phi \over \partial A} = 
    A \left( r - {g^2 \over 2} A^2 \right) \,,
  }
 where we have defined
  \eqn{TauAlphaDef}{
   \tau = 2 p_c^2 + M^2 \qquad 
    \alpha = {g^{7/2} \over \theta^{3/2}} \qquad
    p_c = \sqrt{g \over \theta} \,.
  }
 The various factors of order unity which we have neglected can be
absorbed by rescaling $g$, $p_c$, and $\alpha$.\footnote{In
point of fact, it is difficult to compute some of these factors, since
they arise from the integral equation \OneDimGamma.  We thank
D.~Priour for consultations on the possibility of treating
\OneDimGamma\ numerically.}  None of these
rescalings affects the conclusion that there is a first order
transition.  However, the rest of the analysis is somewhat delicate,
and we must from now on be meticulous about factors.  The calculation
we are about to sketch appeared in \cite{Brazovskii} with some slight
errors, which were corrected in \cite{SwiftH}, modulo a typo in the
sign of the last term in their (2.13).

A stable phase must have a vanishing tadpole.  This happens either
when $A = 0$ (the disordered phase) or when $r - {g^2 \over 2} A^2 =
0$ (the ordered phase).  The respective values of $r$, as well as the
value of $A$ in the ordered phase, can be determined from
  \eqn{rAOrder}{\seqalign{\span\TC}{
   r_o + {\alpha \over \sqrt{r_o}} + \tau = 0 \qquad
   r_d - {\alpha \over \sqrt{r_d}} - \tau = 0 \cr
   A_o = {1 \over g} \sqrt{r_o - {\alpha \over \sqrt{r_o}} - \tau} \,.
  }}
 Note that the ordered phase only exists for $\tau < -3
(\alpha/2)^{2/3} \approx -1.89 \alpha^{2/3}$.  In this range, there
are two solutions for $r_o$ (both positive): let us call these
$r_{o1}$ and $r_{o2}$.  The free energy difference is conveniently
calculated as
  \eqn{PhiDiff}{\eqalign{
   \Delta\Phi &= \Phi_o - \Phi_d = 
     \int_{r_d}^{r_o} dr \, {dA \over dr} {d\Phi \over dA}  \cr
    &= -{g^2 \over 4} A_o^4 + {r_o^2 - r_d^2 \over 2 g^2} + 
     {\alpha \over g^2} (\sqrt{r_o}-\sqrt{r_d})  \cr
    &= {1 \over g^2} \left[
      -{r_o^2 + r_d^2 \over 2} + \alpha (\sqrt{r_o}-\sqrt{r_d}) 
     \right] \,.
  }}
 This equation applies equally to the two solutions for $r_o$.  By
subtracting \PhiDiff\ with $r_o=r_{o1}$ from the same equation with
$r_o=r_{o2}$, one can verify that the larger solution for $r_o$ leads
to the lower free energy.  One can also show, directly from \PhiDiff,
that the disordered phase is favored for $-2.03 \alpha^{2/3} < \tau <
-1.89 \alpha^{2/3}$, and the ordered phase is favored for $\tau <
-2.03 \alpha^{2/3}$.\footnote{In principle one should be able to prove
these assertions using the formula for solutions of a cubic.  However,
since the problem has only one parameter, $\tau/\alpha^{2/3}$, we have
found it more convenient to proceed numerically.}

In terms of the original variables, and again in $d=4$, 
the phase transition to an ordered
phase occurs when $M^2 = -c_1 {g \over \theta} - c_2 {g^{7/3} \over
\theta}$, for some constants $c_1$ and $c_2$ of order unity.  It is
straightforward to show that $c_1 = 1/\pi$ and $\theta p_c^2 =
g/2\pi$.  But to obtain an
accurate value for $c_2$ requires numerics.  The $g^{7/3}$ correction
is a measure of the extent to which the system avoids the second order
behavior expected from a naive one-loop analysis.  Higher loop
corrections are expected to shift the critical value of $M^2 \theta$
by $O(g^3)$.

\subsubsection{Take II}
\label{TakeII}

We now consider a second route to establishing a first order transition
at large $\theta \Lambda^2$ which exploits the intimate connection
between the phases of the noncommutative theory and their limits in
the planar theory. The basic idea is this: As shown earlier, one
cannot have a continuous transition once the disordered phase propagator
has a minimum away from zero which is the case once $\theta \Lambda^2$
is reduced from infinity. While this allows the high temperature solution
to exist below the critical temperature of the planar theory, there
must be a second solution that grows out of the ordered solution of
the planar theory but is now modulated at a very long wavelength.
If we take $\theta \Lambda^2$ sufficiently large at fixed $m^2 < m_c^2$,
the ordered solution will win for it is the correct solution
{\it at} $\theta \Lambda^2 = \infty$. Hence there must be a first order line
emanating from the planar theory critical point.

To construct a proof on these lines it would appear that we could get away
by comparing free energies exactly at $\theta \Lambda^2 = \infty$ for
continuity would give us a range of $\theta \Lambda^2$ over which the
ordering would still hold. This almost works---while there is no problem
with the ordered solution, which connects to the ordered solution of the
planar theory, the delicate point is that the $\theta \Lambda^2 \rightarrow
\infty$ limit of the disordered solution is {\it not} itself a solution of
the planar theory self-consistency equation. Consequently we need to take
the limit carefully, which can be done, modulo a weak and entirely plausible
assumption, without actually solving the equations. (For the simpler
Brazovskii problem, a similar strategy is entirely successful. This is
detailed in the Appendix.)

To begin, let us consider the limiting ordered free energy. This is the
free energy of the planar theory,
\eqn{Ford}{ F_{\rm ord} = 
({m^2 \over 2} \phi_o^2 + {g^2 \over 4} \phi_o^4) - {(m^2 + g^2 \phi_o^2)^2
\over 8 g^2} + {1\over 2} \int {d^d k \over (2\pi)^d} \log[ k^2  +  g^2
\phi_o^2]
}
where $\phi_o = \langle \phi \rangle$ and is determined by $m^2$ via,
\eqn{mVia}{
 m^2 + 4 g^2 \phi_o^2 + 8 g^2 \int {d^d k \over (2\pi)^d} { 1 \over k^2
+  g^2 \phi_o^2} =0 \,.
}
In writing this compact form, we have used the relation $\Sigma = -m^2
+ 4 g^2 \phi_o^2$ valid for the fluctuation propagator in the ordered
phase.

We next turn to the disordered phase self-consistency equation, now written in
terms of the self energy:
  \eqn{SelfGammaReg}{
   \Sigma(p)  = 2 g^2 \int {d^d k \over (2\pi)^d}
     {1 + {1 \over 2} e^{i k \wedge p} \over k^2 + m^2 + \Sigma(k)} \ .
  }
We will assume that the integrals are regulated in the ultraviolet, but the
the precise choice of regulator will not be crucial. Consider the following
assertions:\\
\noindent
(i) At any $\Lambda$ and $\theta \Lambda^2$, $\Sigma(0) = {3\over2}
\Sigma(\infty)$. This follows from the rapid oscillation of the phase factor
as $p \rightarrow \infty$. At increasingly large $\theta \Lambda^2$, this
fall in $\Sigma(p)$ will happen increasingly rapidly. Hence if we examine
$\Gamma^{(2)}(p) = p^2 + m^2 + \Sigma(p)$ it must develop a minimum away from
$p=0$. As already noted this minimum will prevent a continuous transition.

\noindent
(ii) For stability, $\Gamma^{(2)}(p) = k^2 + m^2 + \Sigma(p) > 0$.

\noindent
(iii) It follows that $\Sigma(\infty)$ and $\Sigma(0)$ are positive.

\noindent
(iv) The combination $m^2 + \Sigma(p)$ must change sign between zero
and infinity whenever $m^2 < m_c^2$ and $\theta \Lambda^2$ is
sufficiently large. The proof is by contradiction. At $p=0$ it is positive
by (ii). Were it to remain bounded away from zero we could bound
\eqn{SigmaIneq}{
{1  \over k^2 + m^2 + \Sigma(k)} < {1  \over k^2}
}
and deduce that $\Sigma(\infty) < -m_c^2$ and thence that $\Sigma(\infty)
+ m^2 < m^2 -m_c^2 < 0$ which contradicts the hypothesis.

\noindent
(v) The minimum value of $m^2 + \Sigma(p)$ must vanish as $\theta \Lambda^2
\rightarrow \infty$. This follows from (i) and (ii). The minimum value will
be achieved (or arbitrarily closely approximated in the case of purely
monotone decreasing $\Sigma(p)$) ever closer to $p=0$ in this limit. To
avoid violating (ii), it must vanish.

\noindent
Thus far we have not made any assumptions. Now we need to make a mild
assumption to proceed further.

\noindent
(vi) We assume that $m^2 + \Sigma(\infty)$ vanishes in the large $\theta
\Lambda^2$ limit as well. If $\Sigma(p)$ is monotone decreasing, this
follows from (v). Otherwise is is strongly indicated by continuity from
higher temperatures. For $m^2 > m_c^2$ the limit is a positive constant,
being the renormalized mass of the planar theory, which vanishes at
$m^2 = m_c^2$. It seems highly unlikely that at any fixed $\theta \Lambda^2
\ne \infty$ the large momentum behavior will be non-monotone as a function
of temperature whence the conclusion. In sum, we assume that $\Sigma(p)$
goes pointwise to $-m^2$. Note that this is not a solution of the limit
of \SelfGammaReg.

\noindent
(vii) The remaining task is to evaluate the free energy in this limit.
The free energy of the disordered solution is given by,
\eqn{Fdis}{
 F_{\rm dis} = - {1\over 4} \int {d^d k \over (2\pi)^d} {\Sigma(k) \over
k^2 + m^2 + \Sigma(k)} + {1\over 2} \int {d^d k \over (2\pi)^d}
\log[ k^2 + m^2 + \Sigma(k) ]\ .
}
As the limit of the disordered solution
is not itself a solution, we have to be a bit careful about evaluating
$F_{\rm dis}$ as 
$\theta \Lambda^2 \rightarrow \infty$. Setting $\Sigma(p) + m^2 =0$
in the logarithm is unproblematic. The remaining integral can be evaluated by
breaking it up as
\eqn{SigmaKInt}{\int {d^d k \over (2\pi)^d} {m^2 + \Sigma(k) \over
k^2 + m^2 + \Sigma(k)}
-
\int {d^d k \over (2\pi)^d} {m^2 \over
k^2 + m^2 + \Sigma(k)}
}
wherein the first term vanishes at large $\theta \Lambda^2$ and the second
can be evaluated via the self-consistency equation.
Finally,
\eqn{LimTheta}{ \lim_{\theta \Lambda^2 \rightarrow \infty} F_{\rm dis} =
-{(m^2)^2 \over 8 g^2} + {1\over 2} \int {d^d k \over (2\pi)^d}
\log[ k^2] \ .
}
It is not hard to see that this is higher than the free energy of the
ordered solution for $m^2 < m_c^2$ by expanding the former to leading
order in $\phi_o^2$, which is what we set out to prove.

\subsection{Small $\theta \Lambda^2$: Ising Transition and Lifshitz Point}

We now turn to the opposite limit, namely that in which $\theta
\Lambda^2$ is sufficiently small---we will quantify that below. The
analysis in this limit is straightforward. Exactly at $\theta
\Lambda^2=0$ we recover the commuting theory which has the standard
Ising critical point. At small, non-zero values of $\theta \Lambda^2$
we can expand the exponential phase factor in the quartic vertex in
powers of $\sum_{i<j} p_i \wedge p_j$. The leading term is the quartic
interaction of the commuting theory, and the additional terms are
clearly irrelevant at its critical fixed point in $d \geq 4$. This
statement is true below $d=4$ as well in the $\epsilon$ expansion, for
the additional terms possess a momentum structure that is {\it not}
generated by the interactions already present at the Wilson-Fisher
fixed point.\footnote{This assumes we continue the wedge structure in
some fashion between dimensions four and two.}

This RG statement can be understood more prosaically from the one-loop
computation reviewed in Section~\ref{OneLoopAction}. On examining the 
minimum in $\Gamma^{(2)}(k)$ we find that it continues to be at $k=0$
at small $\theta \Lambda^2$ whence the phase transition would still
be into the uniform Ising symmetry breaking state. As a bonus, we
discover that there is a critical value $ (\theta \Lambda^2)_c \sim 1/g$
at which the minimum starts to move away from $k=0$ and that initially 
it grows as the square root of the deviation in $\theta \Lambda^2$.
While these precise claims will be modified in an actual theory of this
region (which the straight one loop computation is not, being as it is
only valid for $m^2/\Lambda^2 >0$), the general feature that the
Ising transition to a uniform phase will give way to a first order 
transition (expected by continuity with what happens at large 
$\theta \Lambda^2$) to a modulated phase at a Lifshitz point should
be correct. Unfortunately, our self consistent treatment is incapable
of producing a Lifshitz point at a finite temperature ($m^2/\Lambda^2
> -\infty$) in $d=4$ (which is the lower critical dimension in this
approximation) and so the Ising transition and
the first order line meet only at $m^2/\Lambda^2 = -\infty$. Further
investigation of this region is clearly desirable.

\subsection{Odd dimensions}
\label{OddDimensions}

The phase transition we have described is dimension-sensitive.  In
$d=2$, there can be no long-range order: the stripe phase is unstable
to infrared fluctuations.\footnote{This statement is essentially an
application of the Coleman-Mermin-Wagner theorem; however we will
check it explicitly in Section~\ref{StripesSquares}.}  In $d=3$, with
only $\theta^{12} \neq 0$, the story is slightly more complicated,
since $\Gamma^{(2)}(p)$ is no longer a function only of $p^2$.  The
momentum modes for which $\Gamma^{(2)}(p)$ is a minimum lie on a ring
in the $p_1$--$p_2$ plane.  One may nevertheless argue as before that
a second order phase transition is impossible: the integral equation
for $\Gamma^{(2)}(p)$ implies that it is smooth; but then if
$\Gamma^{(2)}(p)$ is to have a zero along the ring, it must be a
quadratic zero, and such a zero renders the integral infinite for all
momenta.  A weakly first order transition to a stripe phase is the
expected behavior (weak because only logarithmic divergences make the
transition first order).  The main point that must be checked is that
the stripe phase is not destroyed by infrared fluctuations.  This we
will do in section~\ref{StripesSquares}.

Suppose we add one more commuting dimension: that is, we work in $d=4$
but with only $\Theta^{12}$ and $\Theta^{21}$ nonzero.
$\Gamma^{(2)}(p)$ should again have a minimum for $p$ on a ring in the
$p_1$--$p_2$ plane.  Since this ring is now codimension three, it is
possible for $\Gamma^{(2)}(p)$ to go smoothly to zero in the
self-consistent Hartree treatment.  So a second order transition
becomes possible.  What happens with a generic $\Theta^{\mu\nu}$ is
less obvious, but again a second order transition seems possible.
One might be able to rule out a second order transition on the basis
of an instability in the one-loop corrected quartic interactions.  It
would be interesting to work out this case in more detail.

\subsection{Ordering beyond Stripes}
\label{StripesSquares}

Our analysis of the ordering in non-commutative scalar theories
has thus far been restricted to stripe phases. Following
Brazovskii, we would expect these to be the correct solution
in the self-consistent Hartree approximation at weak coupling.
However this ignores two orthogonal possibilities. The first
is the option of going between the stripe phase and the 
disordered phase by analogs of the smectic and nematic phases
in liquid crystal physics.\footnote{We should note, however, that in
the extensively studied problem of A--B diblock copolymer melts
\cite{Bates} there appears to be no evidence for intermediate phases.}
 This is conceivable on symmetry
grounds but as it is beyond the reach of our techniques, we
are unable to say anything definite on this score other than
to note that the destruction of long range stripe order
in $d=2$ (see below) would imply that any ordered phase would
necessarily be more symmetric. 

The second possibility is that, at lower temperatures, the
stripe phase might give way to one in which two (or more)
different momentum modes of $\phi$ might condense, leading
to a pattern of squares or rhombi. An interesting question
here is whether the noncommutativity can influence the
state selection in an interesting way.
The purpose of this
section is to argue that stripes are indeed preferred over rhombi for
small $\theta p_c^2$, but that rhombi (or more complicated patterns) 
become favored as $\theta p_c^2$
is increased.  Since $\theta p_c^2$ is small when the coupling $g^2$
is small, the more interesting ordered phases can only arise at strong
coupling.  We will also remark on the first order phase transition to
a triangular crystal that arises when a $\phi^3$ interaction is
present.  The analysis will be very rough in that we will use
a model lagrangian that will incorporate the soft modes and 
nonlinearities in the problem at the tree level instead of attempting
a full fluctuation analysis.  However, the conclusion that stripes 
give way to more complicated patterns as $\theta p_c^2$ is increased
should be robust, since it relies simply on the phases that emerge
from the star product.

As a preliminary, and to introduce our model, let us demonstrate that
long range order for a stripe phase is impossible in two dimensions.
This is a well-established result \cite{Landau}, so we will be brief.
In the case of even dimension $d$, and assuming $\theta^{\mu\nu}$ has
all eigenvalues equal to $\pm\theta$, non-commutativity contributes to
the analysis only by generating a minimum in the propagator for
non-zero momentum $p_c$.  Thus we can work with an effective action of
the Brazovskiian form \cite{Brazovskii}:
  \eqn{SHLreal}{
   S_{\rm eff} = \int d^d x \, \left( 
    {1 \over 2} \kappa_1 \left[ (\partial^2 + p_c^2) \phi \right]^2 +
    {1 \over 2} \kappa_2 \phi^2 + {1 \over 4} \kappa_4 \phi^4 \right) \,.
  }
 A variant for complex scalars is 
  \eqn{SHL}{
   S_{\rm eff} = \int d^d x \, \left(
    \kappa_1 |(\partial^2 + p_c^2) \phi|^2 +
    \kappa_2 |\phi|^2 + 
    {1 \over 2} \kappa_4 |\phi|^4 \right) \,.
  }
 It is assumed that $\kappa_1 > 0$ and $\kappa_4 > 0$.  When $\kappa_2
< 0$, there is pattern formation.  The complex case is simpler, since
the classical minima of \SHL\ are plane waves, $\phi = A e^{i p \cdot
x}$, with $|A| = \sqrt{-\kappa_2/\kappa_4}$ and $|p| = p_c$.  We will
focus on the complex case at the outset, and return to real scalars,
and to $\phi^3$ interactions, near the end of this section.  

The precise claim about absence of long-range order in the model \SHL\
is that at any finite temperature, in $d \leq 3$, long-range order is
destroyed by infrared fluctuations.  To see this, suppose $\phi = A(x)
e^{i p_c x^1}$ where $A(x)$ is assumed to be slowly varying.  Plugging
this ansatz into \SHL, one finds
  \eqn{SHLPlugged}{
   S_{\rm eff} = \int d^d x \, \left(
    \kappa_1 | (2ip_c \partial_{x^1} + \partial_{x^\perp}^2) A |^2 + 
     \kappa_2 |A|^2 + {1 \over 2} \kappa_4 |A|^4 \right) \,.
  }
 The massless modes are those where $A = \sqrt{-\kappa_2/\kappa_4}
e^{i \theta(x)}$.  We will use $x^\perp$ to indicate the directions
perpendicular to the unit vector $\hat{x}^1$.  To quadratic order in
$\theta$, \SHLPlugged\ reduces to
  \eqn{SHLPTwo}{
   S_{\rm eff} = \int d^d x \, \left(
    \left| {\kappa_1 \kappa_2 \over \kappa_4} \right|
     \left[ 4 p_c^2 (\partial_1 \theta)^2 + 
      (\partial_\perp^2 \theta)^2 \right] - 
     {1 \over 2} {\kappa_2^2 \over \kappa_4} \right) \,.
  }
 Because the action is fourth-order in derivatives,
Coleman-Mermin-Wagner arguments are stronger than usual: an estimate
of the fluctuations gives
  \eqn{VEVTheta}{
   \langle \theta^2 \rangle \propto
    \int {d^d q \over 4 p_c^2 q_1^2 + q_\perp^4} \,,
  }
 which is {\it infrared} divergent for $d \leq 3$.

The above argument is not substantially modified in $d=2$ by
non-commutativity because the infrared divergence in \VEVTheta\ refers
to physics at far larger length scales than the scale of
non-commutativity or of frustration.  However, in $d=3$, the premise
of the model is wrong: non-commutativity can only exist for two of the
three coordinates, say $x^1$ and $x^2$.  If a stripe phase forms in
the $x^2$ direction, the dispersion relation (up to dimensionful
constants) is $\omega(q) \sim q_1^4 + q_2^2 + q_3^2$.  The integral
$\int d^3 q / \omega(q)$ is now {\it finite} in the infrared, so the
stripe phase is indeed stable, and our remarks at the end of
Section~\ref{TakeI} were justified.

Let us now proceed to the comparison of energetics for stripes/rolls
and rhombi, in a non-commutative version of the Brazovskii
model,\footnote{We thank E.~Witten for a conversation in which the
simplest case of this argument was worked out for small $\theta
p_c^2$.}
  \eqn{SHNC}{\eqalign{
   S_{\rm eff} &= \int d^d x \, {\cal L}_{\rm eff}  \cr
   {\cal L}_{\rm eff} &= \left(
    \kappa_1 |(\partial^2 + p_c^2) \phi|^2 +
    \kappa_2 |\phi|^2 + 
    {1 \over 2} \kappa_4 \phi\star\phi\star\phi^*\star\phi^* + 
    {1 \over 2} \kappa_{4'} \phi\star\phi^*\star\phi\star\phi^* \right) \,,
  }}
 with $\kappa_1 > 0$, $\kappa_2 < 0$, and $\kappa_4 + \kappa_{4'} >
0$.  For simplicity, we will assume that $\phi$ is modulated at most
in two directions, $x^1$ and $x^2$, which are rotated onto one another
by the action of $\theta^{\mu\nu}$.  More complicated situations can
be imagined since we must assume $d>2$ for the analysis to proceed at
all; however, most of the interesting physics should refer to
dimensions paired together by $\theta^{\mu\nu}$.

Clearly, there are still roll solutions, $\phi = A e^{i p \cdot x}$,
to the equations of motion.  Rhombi would arise from an ansatz $\phi =
A_1 e^{i p_1 \cdot x} + A_2 e^{i p_2 \cdot x}$ with $|p_1| = |p_2| =
p_c$, but with $p_1$ and $p_2$ linearly independent.  This ansatz is
{\it not} a solution to the equation of motion, but if we can show
that it has a lower energy than the roll solution, it is reasonable
evidence that rhombic patterns form.  Plugging the trial function into
\SHNC, we obtain
  \eqn{SHNCAv}{\eqalign{
   \langle |\phi|^2 \rangle &= |A_1|^2 + |A_2|^2  \cr
   \langle \phi\star\phi\star\phi^*\star\phi^* \rangle &=
    |A_1|^4 + |A_2|^4 + 2 (1 + \cos p_1 \wedge p_2) |A_1 A_2|^2  \cr
   \langle \phi\star\phi^*\star\phi\star\phi^* \rangle &=
    |A_1|^4 + |A_2|^4 + 4 |A_1 A_2|^2  \cr
   \langle {\cal L}_{\rm eff} \rangle &= 
    \kappa_2 (|A_1|^2+|A_2|^2) + 
    {1 \over 2} (\kappa_4 + \kappa_{4'}) (|A_1|^2 + |A_2|^2)^2  \cr
    &\quad{} +
     (\kappa_4 \cos p_1 \wedge p_2 + \kappa_{4'}) |A_1 A_2|^2 \,,
  }}
 where $\langle U \rangle$ indicates that $U$ is to be averaged over
space.

Clearly, rolls are favored over rhombi when $\kappa_{4'} >
|\kappa_4|$.  In order for rhombi to be favored over rolls without
making $\kappa_4 + \kappa_{4'} < 0$ (which would destabilize the
theory altogether), we need $\kappa_4 > \kappa_{4'}$ and $p_1 \wedge
p_2$ in the vicinity of $\pi$.  Because
  \eqn{pWedgep}{
   p_1 \wedge p_2 = \theta p_c^2 \sin \angle(p_1,p_2) \,,
  }
 we need $\theta p_c^2 \sim \theta \lambda$ to be finite.  

Suppose we indeed arrange values of the $\kappa_i$ where rhombi are
favored over rolls.  The simplest case is $\kappa_4 > 0$ but
$\kappa_{4'} = 0$.  The curious fact is that a square lattice appears
almost never to be the preferred pattern: if $\theta p_c^2 < \pi/2$,
rolls are preferred; whereas if $\theta p_c^2 > \pi/2$, then among the
trial wave-functions studied so far the one with the lowest energy has
$\angle(p_1,p_2) = \sin^{-1} (2 \theta p_c^2/\pi)$.  We have {\it not}
found actual solutions to the equations of motion with the point
symmetries of a rhombic lattice.  Because of the $\phi^4$ terms, any
such solution would necessarily be a combination of infinitely many
plane waves with wave-vectors on the reciprocal lattice.  It is
straightforward though tedious work to optimize a variational ansatz
which is the sum of finitely many plane waves.  However there is not
much point in going through the exercise because the form of \SHL\ is
only intended to capture the behavior of momenta close to $|p| =
p_c$.

Finally, let us turn to the case of a real scalar.  We will continue
to use the approach of a trial wave-function composed of plane waves
with all momenta on the ring $|p|=p_c$.  Plugging the ansatz
  \eqn{RealTrial}{
   \phi = A_1 e^{i p_1 \cdot x} + A_2 e^{i p_2 \cdot x} + {\rm c.c.} \,,
  }
 into the effective action
  \eqn{RealNC}{\eqalign{
   S_{\rm eff} &= \int d^d x \, {\cal L}_{\rm eff}  \cr
   {\cal L}_{\rm eff} &= 
    {1 \over 2} \kappa_1 \left[(\partial^2 + p_c^2) \phi \right]^2 +
    {1 \over 2} \kappa_2 \phi^2 + 
    {1 \over 4} \kappa_4 \phi\star\phi\star\phi\star\phi \,,
  }}
 with $\kappa_1 > 0$, $\kappa_2 < 0$, and $\kappa_4 > 0$, one obtains
the following:
  \eqn{RealAvs}{\eqalign{
   \langle \phi^2 \rangle &= 2 (|A_1|^2 + |A_2|^2)  \cr
   \langle \phi\star\phi\star\phi\star\phi \rangle &= 
    6 (|A_1|^4 + |A_2|^4) + 
    (16 + 8 \cos p_1 \wedge p_2) |A_1 A_2|^2  \cr
   \langle {\cal L}_{\rm eff} \rangle &= 
    \kappa_2 (|A_1|^2 + |A_2|^2) + 
     {3 \over 2} \kappa_4 (|A_1|^2 + |A_2|^2)^2 + 
     \kappa_4 (1 + 2 \cos p_1 \wedge p_2) |A_1 A_2|^2 \,.
  }}
 The sign on the last term can be negative, and the conclusion is that
on the space of trial wave-functions that we have examined, rhombi are
preferred over stripes if $p_1 \wedge p_2 > 2\pi/3$ is finite.  This
last inequality is possible when $\theta p_c^2 > 2\pi/3$.  In a
similar way, one can consider the condensation of three momentum modes
and show that for $\theta p_c^2$ in a neighborhood of $2\pi/\sqrt{3}$,
a hexagonal pattern is preferred over both rhombi and stripes.  Again,
although these trial wave-functions are not solutions to the equations
of motion, it is reasonable to expect solutions with the same
symmetries to exist and to compete with one another in a similar way.
We have considered condensing momentum modes which all lie in a single
plane on which $\theta_{\mu\nu} = \theta \epsilon_{\mu\nu}$.  In $d=4$
and higher, there are more complicated possibilities where momentum
modes condense in many directions.  We will not attempt to classify
all the possible ordered phases.

One can also consider adding a cubic term to \RealNC:
  \eqn{RealNCMod}{\eqalign{
   S_{\rm eff} &= \int d^d x \, {\cal L}_{\rm eff}  \cr
   {\cal L}_{\rm eff} &= 
    {1 \over 2} \kappa_1 \left[(\partial^2 + p_c^2) \phi \right]^2 +
    {1 \over 2} \kappa_2 \phi^2 + 
    {1 \over 3} \kappa_3 \phi\star\phi\star\phi + 
    {1 \over 4} \kappa_4 \phi\star\phi\star\phi\star\phi \,.
  }}
 As explained in \cite{Seiberg}, a $\phi^{\star 3}$ interaction
contributes a term $-I_2(p)$ to $\Gamma^{(2)}(p)$; thus if the cubic
interaction is strong as compared to the quartic interaction,
$\Gamma^{(2)}(p)$ is monotonic, and there is no reason to think that
there are spatially non-uniform phases at all.  For the purposes of
this discussion, let us assume then that the cubic interaction is weak
but nonzero.

Given that the leading nonlinearity is cubic, the natural expectation
in two dimensions would
be that commutative versions of \RealNCMod\ exhibit a first order transition 
to hexagonal crystals as $\kappa_2$ is lowered.  This persists in the presence of
non-commutativity, as the following computation shows.  The ansatz for
$\phi$ is
  \eqn{phiAnsatz}{
   \phi = A_1 e^{i p_1 \cdot x} + A_2 e^{i p_2 \cdot x} + 
    A_3 e^{-i (p_1+p_2) \cdot x} + {\rm c.c.} \,,
  }
 with $|p_1| = |p_2| = |p_1+p_2| = p_c$.  In two dimensions, the only
way that this can be arranged is for $p_1$, $p_2$, and
$-p_1-p_2$ to point to the vertices of an equilateral triangle
centered on the origin.  The relevant spatial averages are
  \eqn{CubicAv}{\eqalign{
   \langle \phi^2 \rangle &= 2 \sum_i |A_i|^2  \cr
   \langle \phi\star\phi\star\phi \rangle &= 3 \cos {p_1 \wedge p_2 \over 2}
    (A_1 A_2 A_3 + A_1^* A_2^* A_3^*)  \cr
   \langle \phi\star\phi\star\phi\star\phi \rangle &= 
    6 \left( \sum_i |A_i|^2 \right)^2 + 
    4 (1 + 2 \cos p_1 \wedge p_2) \sum_{i<j} |A_i A_j|^2  \cr
   \langle {\cal L}_{\rm eff} \rangle &= 
    \kappa_2 \sum_i |A_i|^2 + \kappa_3 \cos {p_1 \wedge p_2 \over 2}
     (A_1 A_2 A_3 + A_1^* A_2^* A_3^*)  \cr
    &\quad{} + 
    {3 \over 2} \kappa_4 \left( \sum_i |A_i|^2 \right)^2 +
    \kappa_4 (1 + 2 \cos p_1 \wedge p_2) \sum_{i<j} |A_i A_j|^2 \,.
  }}
 As long as the cubic term is present, a first order transition to a
hexagonal lattice is the expected behavior.  To be more precise, the
expectation is that with $\kappa_2$ sufficiently small but positive,
there is a minimum where $\langle {\cal L}_{\rm eff} \rangle$ is
negative, and where all three $A_i$ are nonzero.  To see that this
must be so, set $A_i = A n_i$ with $\sum_i n_i^2 = 1$.  Then
  \eqn{TildeAv}{\eqalign{
   \langle {\cal L}_{\rm eff} \rangle &= \tilde\kappa_2 A^2 + 
    \tilde\kappa_3 A^3 + \tilde\kappa_4 A^4  \cr
   \tilde\kappa_2 &= \kappa_2  \cr
   \tilde\kappa_3 &= 2 \kappa_3 \cos {p_1 \wedge p_2 \over 2} 
    n_1 n_2 n_3  \cr
   \tilde\kappa_4 &= \kappa_4 \left[ {3 \over 2} +
     (1 + 2 \cos p_1 \wedge p_2) \sum_{i<j} n_i^2 n_j^2 \right] \,.
  }}
 One may easily show that $\tilde\kappa_4 > 0$, so there is no runaway
behavior.  It is straightforward to show that $\langle {\cal L}_{\rm
eff} \rangle$ attains negative values precisely if
$\sqrt{\tilde\kappa_2 \tilde\kappa_4} < 2 |\tilde\kappa_3|$.  This is
the desired result: even assuming a small cubic interaction, the first
order transition at finite positive $\kappa_2$ swamps the would-be
second order behavior at $\kappa_2 = 0$.  The only exception is when
$\cos {p_1 \wedge p_2 \over 2} = 0$: at this point, the cubic term
vanishes.  A solution with $|A_1| = |A_2| = |A_3|$ is still preferred,
but the phase of $A_1 A_2 A_3$ is a flat direction.  On general
Landau-Ginzburg theory grounds one might expect a second order point
at $\cos {p_1 \wedge p_2 \over 2} = 0$ separating the two first order
lines corresponding, respectively, to phase locking of the three plane
waves with $A_1 A_2 A_3$ positive or negative; however it seems likely
that fluctuations again drive the transition first order.  In higher
dimensions the story is again more complicated.  The usual expectation
is a lattice with the maximal number of equilateral triangles; in the
presence of non-commutativity these lattices are likely to have
deformations and preferred orientations.  We leave an investigation of
the possibilities for future work.

\section{The non-commutative $O(N)$ vector model}
\label{ONVector}

Naively, there are two reasons to think that quantum field theories
defined on non-commutative spaces are under better control
perturbatively than their commutative counterparts.  First, non-planar
loop diagrams include an oscillatory factor in the integrand which
generically cures ultraviolet divergences \cite{Seiberg}.  Second,
compactifying the non-commutative position space also entails a
compactification in momentum space \cite{AmbjornOne}: this is one of
several manifestations of the interplay between ultraviolet and
infrared effects.  But there are also reasons to think that such
quantum field theories are not under good perturbative control.  First
and perhaps most seriously, it has not been shown that the special
form of the tree level lagrangian (polynomial in derivatives and
fields, but with all ordinary products replaced by star products) is
preserved by quantum corrections beyond one loop.  Second, planar
divergences are just as bad as for commutative field theories.

The latter two difficulties can be resolved for special quantum field
theories.  Consider the $O(N)$ vector model:
  \eqn{ONActAgain}{
   S = \int d^d x \, \left[ {1 \over 2} (\partial\phi_i)^2 + 
    {1 \over 2} m^2 \phi_i^2 + {g^2 \over 4} \phi_i \star \phi_i 
     \star \phi_j \star \phi_j + 
    {g'^2 \over 4} \delta^{ik} \delta^{jl} 
      \phi_i \star \phi_j \star \phi_k \star \phi_l \right] \,,
  }
 which, in the large $N$ limit, is dominated by bubble graphs in its
disordered phase.  If
$g^2 = 0$, the loop integrals in these graphs converge even without an
ultraviolet cutoff, due to the oscillatory factor in the loop
integrands.  Thus there are {\it no counterterms} in the large $N$
limit: the theory is perfectly finite.  At subleading orders in $N$,
counterterms do arise, and all the usual questions arise regarding
whether the special form of the lagrangian is preserved in the quantum
theory.  However it can perhaps be regarded as progress toward
deciding the issue of perturbative renormalizability that these
difficulties can be suppressed by $1/N$ in an appropriate 't~Hooftian
limit.

Returning to the bubble sum for \ONActAgain\ with $g^2=0$, this can
be reduced to an integral equation---the 1PI two-point function is given
implicitly by
  \eqn{Bubbles}{
   \Gamma^{(2)}(p) = p^2 + m^2 +
    \int {d^d k \over (2\pi)^d} {g'^2 N e^{i p \wedge k}
    \over \Gamma^{(2)}(k)} \,.
  }
 In other words, the self-consistent Hartree approximation in the
disordered phase is exact at large $N$. The existence of an ordered
phase, following our earlier caveat, remains an open question. 

It may be that \ONActAgain\ with $g^2 = 0$ is a special case of a more
general strategy for generating quantum field theories with improved
convergence properties in a special limit---the main ingredient being
two conflicting notions of planarity, one from the index structure of
the fields, and one from the star product.  However, we are {\it not}
guaranteed to obtain a finite quantum field theory (even in special
limits) through this trick, as there may be graphs which are planar in
both senses.  For example, the lagrangian
  \eqn{MatrixL}{
   {\cal L} = {1 \over 2} \tr (\partial\Phi)^2 + {1 \over 2} m^2 \tr \Phi^2
    + {g'^2 \over 4} \delta^{i_2 k_1} \delta^{k_2 j_1} \delta^{j_2 l_1}
      \delta^{l_2 i_1} \Phi_{i_1 i_2} \star \Phi_{j_1 j_2} \star
      \Phi_{k_1 k_2} \star \Phi_{l_1 l_2} \,,
  }
 where $\Phi$ is an $N \times N$ hermitian matrix, specifies a quantum
field theory with two conflicting notions of planarity; but there are
still certain planar diagrams diverge, for instance the one-loop
correction to the quartic interaction vertex.  This divergence
requires a counterterm of the form $\tr
(\Phi\star\Phi\star\Phi\star\Phi)$, so we learn that the special form
of the lagrangian \MatrixL\ is {\it not} preserved by quantum
corrections.

\section{Continuum limits}
\label{Continuum}

We turn now the question of taking a continuum limit, i.e.\ the
question of renormalizability. We will examine this in the disordered
phase of the theory within our self-consistent Hartree
approach. Formally, we will consider the renormalization of the
infinite $N$ theory, \ONActAgain, with $g^2 N$ and $g'^2 N$ finite.
In this approximation, the only non-trivial correlation function that
enters the theory is $\Gamma^{(2)}(p)$ but on account of the
noncommutativity it has significant momentum dependence that makes the
procedure somewhat more complicated than it might seem at first sight.
Nevertheless, the claim is that there are non-trivial continuum
limits, and therefore interacting renormalizable field theories, in
any even dimension.  It was conjectured in \cite{Seiberg} that
non-commutative $\phi^4$ theory should be renormalizable in $d=4$,
despite the infrared divergences.  The results of this section do not
amount to a demonstration of this claim, because we persist in working
at infinite $N$; however we hope that an extension to finite $N$ may
be possible. We should note though, that our Wilsonian attempt to
renormalize by means of the planar theory critical point is closely
connected to the the planar subtraction algorithm suggested in 
\cite{Seiberg}.  

As a warmup recall the problem of the commuting, infinite $N$, $O(N)$
vector model. Here,
\eqn{GammaFromSigma}{ \Gamma^{(2)}(p) = p^2 + m^2 + \Sigma}
where
\eqn{SigmaG}{
 \Sigma = {g^2 \over 2} \int^\Lambda {d^d k \over (2\pi)^d} {1
    \over k^2 + m^2 + \Sigma} \ ,
}
where we have indicated the regulation of the integral explicitly. In $2<d<4$,
this theory has a critical point $m^2 = m_c^2(\Lambda)$ at which the 
renormalized mass $m^2 + \Sigma$ vanishes. Explicitly, 
\eqn{mcGiven}{m_c^2 = - {g^2 \over 2}
\int^\Lambda {d^d k \over (2\pi)^d} {1 \over k^2} \ ,
}
but that is not central. 

From the existence of the critical point it follows that we may choose
$m^2(\Lambda)$ in order that $m^2(\Lambda) + \Sigma(g^2,\Lambda)=M^2$
is held fixed,\footnote{This is the point in the argument where the 
existence of a critical point in the bare theory is crucial. As we
know that we can tune the bare mass to get $M^2$ to vanish at any
value of the cutoff, it follows that we can hold it fixed at a specified 
value as well.} whereupon we may take $\Lambda$
to infinity and obtain a renormalized propagator $ \Gamma_R^{(2)}(p) =
p^2 + M^2$. In this example the end product of this mass
renormalization is trivial, but the underlying lattice problem is
not. A non-trivial correlation length exponent $\nu$ is hidden in the
relation between the bare mass and the renormalized mass.

Turning now to the noncommuting infinite $N$ theory, we can divide 
$\Sigma(p)$ into the parts coming from the planar and non-planar
diagrams and rewrite the self consistency equation in two parts:
\eqn{PlanarPart}{
M^2 = m^2 + 2 g^2 N \int^\Lambda {d^d k \over (2\pi)^d} {1
    \over k^2 + M^2 + \Sigma_{\rm np}(k)}
}
and
  \eqn{SigmaNP}{
\Sigma_{\rm np}(p) = g'^2 N 
\int^\Lambda {d^d k \over (2\pi)^d} {
e^{i k \wedge p} \over k^2 + M^2 + \Sigma_{\rm np}(k)} \ .
  }
Note that $M^2 - m^2 \equiv \Sigma(\infty)$ introduced earlier.

Now, the planar theory has a critical point at $m^2 = m_c^2$ and 
$\theta \Lambda^2 = \infty$ where $M^2$ vanishes. Keeping a fixed dimensionful 
$\theta$ then sends $\theta \Lambda^2$ to infinity automatically,
while the critical point enables us to choose $m^2(\Lambda;\theta)$ such
that $M^2$ is held fixed as $\Lambda \rightarrow \infty$.

In this limit we are able to remove the cutoff from \SigmaNP\ as well
which still defines a finite $\Sigma_{\rm np}(p)$ and thereby a finite
renormalized two point function $ \Gamma_R^{(2)}(p) = p^2 + M^2 +
\Sigma_{\rm np}(p)$. Note that by dimensional analysis $\Sigma_{\rm
np}$ has the scaling form $\Sigma_{\rm np} = M^2 f(p/M, \theta
M^2)$. Note also, that $\Sigma_{\rm np} \sim {1\over (\theta p)^{d-2}}$ is
divergent in the infrared.\footnote{Note that this is the
{\it same} integral equation that arose in \Bubbles, albeit with $M^2$
replacing $m^2$.}

In this fashion we see that the existence of the planar theory
critical point does allow a continuum limit to be taken at
$N=\infty$. There is one more limiting theory in this case,
the massless purely planar theory obtained by sitting exactly
at the critical point. (If the $N=\infty$ theory has a 
broken symmetry phase then a third continuum limit would be
feasible from within that phase.)
We expect that the planar limit does not change much between
$N=\infty$ and $N=1$ in $d \geq 4$. That suggests that more progress
could be made in establishing renormalizability via an $1/N$ 
expansion---the same logic underlies the comparable demonstration
for commutative scalar theories. However, we have not examined
this question in any detail.

Note also the interesting feature that the continuum limit is
nontrivial in the infrared while the ultraviolet behavior is free. The
latter is consistent with the triviality of commuting scalar theories
in $d \geq 4$ although in this case it cannot be distinguished from
the vanishing of the anomalous dimension of the scalar field at
$N=\infty$. At any rate, the point is that the continuum limit will be
nontrivial even above $d=4$ (say in $d=6$) for finite noncommutativity
is, in a sense, a relevant perturbation at the planar theory fixed
point. This is in contrast to the situation in the commuting case
where only the free massive theory is possible.

Finally, we should note that a different set of continuum limits is
possible near the Lifshitz point in the $N=1$ theory. These would entail 
keeping $\theta \Lambda^2$ finite as $\Lambda \rightarrow \infty$ and hence 
a vanishing $\theta$. We have not studied these, but our large $N$
approach could be extended above $d=4$ should interest in these theories
be warranted.

\noindent
\section{Concluding Remarks}
\label{Concluding}

The purpose of this paper has been to investigate unusual phase
structure in the simplest interacting non-commutative field theory,
namely $\phi^4$ theory.  We expect the existence of stripe phases to
be quite common in non-commutative theories, the reason being that
$\Gamma^{(2)}(p)$ typically has singular behavior as $p \to 0$ when
the cutoff is removed.  If $\Gamma^{(2)}(p) \to -\infty$ as $p \to 0$,
the theory is sick in the sense that for no value of bare masses have
we found a stable vacuum.  If $\Gamma^{(2)}(p) \to +\infty$, then some
version of the arguments of Section~\ref{Phase} should establish a
first order transition to a stripe phase.

It is natural to ask, what manifestation might this transition find in
string theory?  At present, we have no definite answer; but let us
remark on one obvious venue where such a transition might be expected
to arise.  D-branes in bosonic string theory can, in a rough
approximation, be thought of as classical lumps of an open string
tachyon field $T$ (see for example \cite{HarveyKraus}).  The potential
for the tachyon field is cubic, so non-commutativity (in the form of a
$B_{\mu\nu}$-field) merely destabilizes the $T=0$ vacuum further.
Unstable D-branes in type II superstrings, however, have a quartic
potential for the tachyon field, and it is possible that a stripe
phase will arise for sufficiently large $B_{\mu\nu}$.  If a stripe
phase indeed exists, it would be quite a peculiar string background,
not readily comprehensible in classical terms.  Classically speaking,
the stripe phase would amount to an alternating arrangement of stable
D-branes and their anti-branes, parallel to one another.  Such an
arrangement seems obviously unstable toward the branes and anti-branes
collapsing into one another.  There are however some caveats: first,
in working with the non-commutative field theory as an approximation
to the string dynamics, we are by assumption working in a limit where
the closed string interactions are suppressed.  So the collapse of
branes into anti-branes could have a much longer time scale than the
transition to a stripe phase from a $T=0$ phase.  Second, working
directly with a field theory is suspect because the cutoff
($1/\sqrt{\alpha'}$) is on the same order as the tachyon mass.  A full
string theory computation, with finite $\alpha'$, seems to be the only
wholly reliable approach.

We should note that our description of stripe phases may not
incorporate an important piece of the physics of unstable D-branes:
namely that the open string degrees of freedom are believed to be
confined when the tachyon field has condensed.  The mechanism for this
confinement is not yet wholly clear (see \cite{Yi:1999hd,Sen:1999md}
for two interesting proposals), but it could play a role in the
stability of various phases.

A relatively well-understood aspect of unstable D-branes in a strong
background $B_{\mu\nu}$ field is condensation into a semiclassical
configuration known as the non-commutative soliton \cite{GMS}.  It has
been argued \cite{Harvey} that this configuration represents the
collapse of a bosonic Dp-brane into a D(p-2)-brane; see also
\cite{Mukhi} for closely related work.  The role of these objects in
our quantum considerations is not entirely clear. For one thing, the
stable solitons/instantons at infinite $\theta$ found in \cite{GMS} by
leaving out the gradient term in the action, have exactly zero action
in our case which makes it imperative to keep the gradient term in the
analysis. Should such an improved computation be feasible, we would
still suspect that in the large $N$ limit these objects will be
unstable to unwinding via other directions in field space and hence do
not play a role.

In the $N=1$ case, the solitons/instantons {\it could} exist. As
instantons we would have to consider whether they invalidate
our conclusions on the nature of the ordered phases. This seems
unlikely, at least at weak coupling where the wavelength of our stripe
phases is much larger, by a factor of $1/\sqrt{g}$, than the size
of the instantons in the classical analysis. More generally, they
are local distortions of the condensate, but do not (unlike, say
vortices in an XY model) affect the ordering at large distances.
Hence we expect that they will renormalize the properties of the
ordered phases but not lead to any singular effects. 

In contrast, actual solitons in a quantum theory in odd dimensions
would rely on the existence of an ordered phase. In the classical
analysis, this is assumed to be a uniform condensate but as
we have seen this is no longer the case in the quantum theory. Again,
at weak coupling, it is plausible that the solitons are essentially
undistorted on account of their size, but their dynamics would
clearly be affected by motion in an inhomogeneous background.

It is perhaps worth recalling some features of quantum Hall physics
that may find a formulation as noncommutative field theories. The
first is the physics of the $\nu =1/2$ state which has an elegant
microscopic interpretation in terms of dipoles\footnote{For a
recent discussion including an invocation of non-commutative
geometry, see \cite{Read98}.} with evident parallels to 
the discussion of Bigatti and Susskind \cite{SusskindBigatti}. 
We note that stripe phases do occur in quantum Hall systems
at half filling (in high Landau levels) \cite{Koulakov,Moessner}
but hasten to add that any connection between them, the dipolar
$\nu=1/2$ theory and our noncommutative results is purely speculative
at this point.

An equally speculative connection is to a widely studied model of the integer 
quantum Hall effect, namely 
Pruisken's non-linear sigma model.  It has several incarnations, but
the simplest has a lagrangian of the form
  \eqn{Pruisken}{
   {\cal L} = {1 \over 4} \sigma_{xx}^{(0)} \tr (\partial T)^2
    + {1 \over 8} \sigma_{xy}^{(0)} \tr \left( T 
     [\partial_x T,\partial_y T] \right) \,,
  }
 where $T$ takes values in the coset $U(2N)/(U(N) \times U(N))$.  The
second term is topological.  The model arises from a replica treatment
of disorder in a theory of non-interacting electrons in a magnetic
field.  In the end, physical quantities must be computed in a $N \to
0$ limit.  In the original formulation, the RG flow in the 
$(\sigma_{xy},\sigma_{xx})$ plane was argued
to have fixed points at $\sigma_{xx} = 0$ and $\sigma_{xy}
\in {\bf Z}$ and at $\sigma_{xx} = 1/2$ and $\sigma_{xy} \in {\bf Z} +
1/2$.\footnote{However, Zirnbauer \cite{Zirnbauer} has argued persuasively
that the fixed point theories have instead a Wess-Zumino-Witten form.}  
The first set of fixed points represent the quantum Hall
plateaux, and the second set control the critical behavior of
transitions between plateaux.  The behavior $\sigma_{xx} \to 0$ in the
infrared is the analog of confinement in this model, and the special
fixed points at non-zero $\sigma_{xx}$ are believed to arise from
instanton effects, similar to 't~Hooft's treatment of deconfinement in
QCD at $\theta = \pi$.

It is suggestive that precisely the $U(2N)/(U(N) \times U(N))$ coset
arises as the vacuum manifold in the condensation of unstable D-branes
in type II string theory.  However it is difficult to see how the
topological term $\tr \left( T [\partial_x T,\partial_y T] \right)$
can arise in the string effective action.  This term is the leading
order term in a derivative expansion of $\tr \left( T^{\star 3} - T^3
\right)$; however, from a type II string theory point of view, neither
the mixing of star products with ordinary products inside a single
trace, nor the cubic form of this potential, is expected.
Nevertheless, strictly from a field theory point of view, it would be
interesting to ask whether replacing the topological term with $\tr
\left( T^{\star 3} - T^3 \right)$ preserves the universality class.
We hope to return to these issues in the future.

\section*{Acknowledgements}

We would like to thank D.~Huse for many useful discussions and for pointers 
to the literature on the Brazovskii transition.  The research of
S.S.G.\ was supported in part by DOE grant~DE-FG02-91ER40671 while S.L.S.
is grateful to the NSF (grant~DMR 99-7804), the Alfred P.~Sloan 
Foundation and the David and Lucille Packard Foundation for support.

\section*{Appendix: The Brazovskii Transition by Non-Bra\-zov\-skiian Means}

Here we will give a version of the argument used in
Section~\ref{TakeII} for the transition studied originally by
Brazovskii \cite{Brazovskii}. The relevant action is \SHLreal, which
we rewrite as
  \eqn{SH}{\eqalign{
   S_{\rm eff} &= \int d^d x \, {\cal L}_{\rm eff}  \cr
   {\cal L}_{\rm eff} &= 
    {\kappa \over 2}((\partial^2 + p_c^2) \phi)^2 + {1 \over 2} m^2 \phi^2
    + {g^2 \over 4!} \phi^4 \ .
  }}
The traditional route to showing that there is a first order transition
is a weak coupling analysis. Here we will work at fixed coupling but 
instead take $p_c$ to be the ``small parameter''. For this to work we
will need that the theory at $p_c=0$ possess a continuous transition to
a broken symmetry phase, else nothing is gained. In the self-consistent
Hartree approximation of Brazovskii for the high temperature phase,
  \eqn{SelfGammaBr}{
   \Sigma = 
    {g^2 \over 2} \int {d^d k \over (2\pi)^d} 
     {1  \over \kappa (k^2 - p_c^2)^2 + m^2 + \Sigma}
  }
this requires that we take $d > 4$.\footnote{Alternatively, we can
modify the Brazovskii action by replacing $(k^2 - p_c^2)^2$ by
$ (k-p_c)^2$ which preserves the feature of a codimension one
surface of soft modes while retaining the connection to a critical
point in $d >2$.} In
these dimensions, the critical point of the uniform ($p_c=0$) theory
is at $m_c^2 = - {g^2 \over 2} \int {d^d k \over (2\pi)^d} {1 \over
\kappa k^4}$.

Now consider $p_c >0$ at fixed $g^2$. By the standard argument
sketched in Section~\ref{TakeI}, there is always a disordered solution
at any $m^2/\Lambda^2 > -\infty$, with a free energy (written in terms
of self-consistent parameters)
\eqn{DisFreeBr}{
F = -{\Sigma^2 \over 2 g^2} + {1\over 2} \int {d^d k \over (2\pi)^d}
\log[ \kappa (k^2 - p_c^2)^2 + m^2 + \Sigma] \ ,
}
whence a continuous transition out of the high temperature phase is impossible.

Next, we follow the disordered solution as we take $p_c$ to zero. For $m^2 >
m_c^2$ the solution smoothly connects to the disordered solution of the uniform
theory with $\Sigma + m^2 >0$. But for $m^2 < m_c^2$ this is not possible as
there is only an ordered solution. So in the vicinity of the $p_c=0$ line we
must have two solutions, where the second evolves out of the broken symmetry
solution of the uniform theory as $p_c$ is tuned away from zero. 

\begin{figure}
   \centerline{\psfig{figure=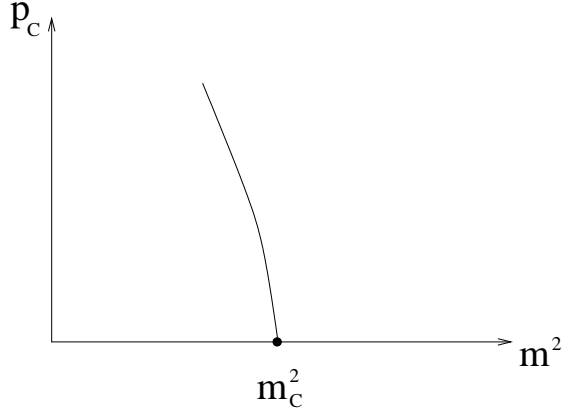,width=2.9in}}
   \caption{A first order line ending at a second order point.}\label{figH}
  \end{figure}

The remaining task is to show that for $m^2 < m_c^2$ there is always a value of
$p_c$ below which the ordered solution wins---this then shows, as in the example
of the noncommutative theory in Section~\ref{Phase},  that there is a first order line 
in the $m^2,p_c$ plane (see Figure~\ref{figH}). For this again it suffices to make the 
comparison in the limit $p_c \rightarrow 0$ on the grounds that free energies are 
continuous.

The disordered solution is readily shown to require that $m^2 + \Sigma \rightarrow 
0$ as $p_c \rightarrow 0$ whenever $m^2 < m_c^2$. Hence its free energy has the
limit,
\eqn{FLimit}{ 
F = -{(m^2)^2 \over 2 g^2} + {1\over 2} \int {d^d k \over (2\pi)^d}
\log[ \kappa k^4 ] \ .
}
The free energy of the ordered solution at $p_c=0$ is
\eqn{FordPc}{
 F = ({m^2 \over 2} \phi_o^2 + {g^2 \over 4!} \phi_o^4) - 
{(m^2 + {g^2 \over 6}  \phi_o^2)^2
\over 2 g^2} + {1\over 2} \int {d^d k \over (2\pi)^d} \log[ \kappa
 k^4  + {g^2\over3}
\phi_o^2]
}
where the parameters satisfy,
\eqn{ParamSatisfy}{
 m^2 + {g^2 \over 6} \phi_o^2 + {g^2 \over 2} \int {d^d k \over (2\pi)^d} 
{ 1 \over \kappa k^4 + {g^2 \over 3} \phi_o^2} =0 }
and
\eqn{SigmaSatisfy}{ \Sigma = {g^2 \over 3} \phi_o^2 \ .}

An explicit comparison, easily done perturbatively in $\phi_o^2$, shows that
the ordered solution has lower energy, which is what we set out to show.

\bibliography{phase}
\bibliographystyle{ssg}

\end{document}

%% file: lmacros.tex
\input umacros
\def\lbldef#1#2{\expandafter\gdef\csname #1\endcsname {#2}}
\def\eqn#1#2{\lbldef{#1}{(\ref{#1})}%
\begin{equation} #2 \label{#1} \end{equation}}
\def\eqalign#1{\vcenter{\openup1\jot
    \halign{\strut\span\TL & \span\TR\cr #1 \cr
   }}}
\def\eno#1{(\ref{#1})}

%% file: umacros.tex
\def\TL{\hfil$\displaystyle{##}$}
\def\TR{$\displaystyle{{}##}$\hfil}
\def\TC{\hfil$\displaystyle{##}$\hfil}
\def\TT{\hbox{##}}
\def\seqalign#1#2{\vcenter{\openup1\jot
  \halign{\strut #1\cr #2 \cr}}}

\def\comment#1{}
\def\fixit#1{}

\def\mop#1{\mathop{\rm #1}\nolimits}

\def\tr{\mop{tr}}

\def\overleftrightarrow#1{\vbox{\ialign{##\crcr
     $\leftrightarrow$\crcr\noalign{\kern-0pt\nointerlineskip}
     $\hfil\displaystyle{#1}\hfil$\crcr}}}

\def\lsim{\mathrel{\mathstrut\smash{\ooalign{\raise2.5pt\hbox{$<$}\cr\lower2.5pt\hbox{$\sim$}}}}}
\def\gsim{\mathrel{\mathstrut\smash{\ooalign{\raise2.5pt\hbox{$>$}\cr\lower2.5pt\hbox{$\sim$}}}}}

\def\sqr#1#2{{\vcenter{\vbox{\hrule height.#2pt
         \hbox{\vrule width.#2pt height#1pt \kern#1pt
            \vrule width.#2pt}
         \hrule height.#2pt}}}}

\def\href#1#2{#2}  


%% file: phase.bbl
\begingroup\raggedright\begin{thebibliography}{10}

\bibitem{Filk}
T.~Filk, ``Divergencies in a field theory on quantum space,'' {\em Phys. Lett.}
  {\bf B376} (1996) 53.

\bibitem{cds}
A.~Connes, M.~R. Douglas, and A.~Schwarz, ``Noncommutative geometry and matrix
  theory: Compactification on tori,'' {\em JHEP} {\bf 02} (1998) 003,
  \href{http://xxx.lanl.gov/abs/hep-th/9711162}{{\tt hep-th/9711162}}.

\bibitem{SusskindBigatti}
D.~Bigatti and L.~Susskind, ``Magnetic fields, branes and noncommutative
  geometry,'' \href{http://xxx.lanl.gov/abs/hep-th/9908056}{{\tt
  hep-th/9908056}}.

\bibitem{SeibergWitten}
N.~Seiberg and E.~Witten, ``String theory and noncommutative geometry,'' {\em
  JHEP} {\bf 09} (1999) 032, \href{http://xxx.lanl.gov/abs/hep-th/9908142}{{\tt
  hep-th/9908142}}.

\bibitem{ChepelevRoiban}
I.~Chepelev and R.~Roiban, ``Renormalization of quantum field theories on
  noncommutative R**d. I: Scalars,''
  \href{http://xxx.lanl.gov/abs/hep-th/9911098}{{\tt hep-th/9911098}}.

\bibitem{Seiberg}
S.~Minwalla, M.~V. Raamsdonk, and N.~Seiberg, ``Noncommutative perturbative
  dynamics,'' \href{http://xxx.lanl.gov/abs/hep-th/9912072}{{\tt
  hep-th/9912072}}.

\bibitem{GMS}
R.~Gopakumar, S.~Minwalla, and A.~Strominger, ``Noncommutative solitons,'' {\em
  JHEP} {\bf 05} (2000) 020, \href{http://xxx.lanl.gov/abs/hep-th/0003160}{{\tt
  hep-th/0003160}}.

\bibitem{Harvey}
J.~A. Harvey, P.~Kraus, F.~Larsen, and E.~J. Martinec, ``D-branes and strings
  as non-commutative solitons,''
  \href{http://xxx.lanl.gov/abs/hep-th/0005031}{{\tt hep-th/0005031}}.

\bibitem{Rivelles}
H.~O. Girotti, M.~Gomes, V.~O. Rivelles, and A.~J. da~Silva, ``A consistent
  noncommutative field theory: The Wess-Zumino model,''
  \href{http://xxx.lanl.gov/abs/hep-th/0005272}{{\tt hep-th/0005272}}.

\bibitem{Campbell}
B.~A. Campbell and K.~Kaminsky, ``Noncommutative field theory and spontaneous
  symmetry breaking,'' {\em Nucl. Phys.} {\bf B581} (2000) 240--256,
  \href{http://xxx.lanl.gov/abs/hep-th/0003137}{{\tt hep-th/0003137}}.

\bibitem{Moore}
M.~A. Moore, T.~J. Newman, A.~J. Bray, and S.-K. Chin {\em Phys. Rev.} {\bf
  B58} (1998) 936.

\bibitem{Golubovic}
L.~Golubovic, ``Phase transitions and the ordered state in models with a
  continuous set of energy minima in large-$N$ limit,'' {\em Phys. Rev.} {\bf
  B27} (1983) 4488--4490.

\bibitem{Fischler:2000fv}
W.~Fischler {\em et.~al.}, ``Evidence for winding states in noncommutative
  quantum field theory,'' {\em JHEP} {\bf 05} (2000) 024,
  \href{http://xxx.lanl.gov/abs/hep-th/0002067}{{\tt hep-th/0002067}}.

\bibitem{Fischler:2000bp}
W.~Fischler {\em et.~al.}, ``The interplay between theta and T,'' {\em JHEP}
  {\bf 06} (2000) 032, \href{http://xxx.lanl.gov/abs/hep-th/0003216}{{\tt
  hep-th/0003216}}.

\bibitem{Arcioni:2000bz}
G.~Arcioni, J.~L.~F. Barbon, J.~Gomis, and M.~A. Vazquez-Mozo, ``On the stringy
  nature of winding modes in noncommutative thermal field theories,''
  \href{http://xxx.lanl.gov/abs/hep-th/0004080}{{\tt hep-th/0004080}}.

\bibitem{Chaikin}
P.~M. Chaikin and T.~C. Lubensky, {\em Principles of Condensed Matter Physics}.
\newblock Cambridge University Press, Cambridge, 1995.

\bibitem{Polchinski84}
J.~Polchinski, ``Renormalization and effective lagrangians,'' {\em Nucl. Phys.}
  {\bf B231} (1984) 269--295.

\bibitem{Girvin86}
S.~M. Girvin, A.~H. MacDonald, and P.~M. Platzman, ``Magneto-roton theory of
  collective excitations in the fractional quantum Hall effect,'' {\em Phys.
  Rev.} {\bf B33} (1986) 2481.

\bibitem{ArefOne}
I.~Y. Aref'eva, D.~M. Belov, and A.~S. Koshelev, ``A note on UV/IR for
  noncommutative complex scalar field,''
  \href{http://xxx.lanl.gov/abs/hep-th/0001215}{{\tt hep-th/0001215}}.

\bibitem{GrossMiljkovic}
D.~J. Gross and N.~Miljkovic, ``A nonperturbative solution of d=1 string
  theory,'' {\em Phys. Lett.} {\bf B238} (1990) 217.

\bibitem{Nishigaki}
S.~Nishigaki, ``Wilsonian approximated renormalization group for matrix and
  vector models in $2<d<4$,'' {\em Phys. Lett.} {\bf B376} (1996) 73--81,
  \href{http://xxx.lanl.gov/abs/hep-th/9601043}{{\tt hep-th/9601043}}.

\bibitem{Ferretti}
G. Ferretti, unpublished.

\bibitem{Brazovskii}
S.~A. Brazovkii, ``Phase transition of an isotropic system to a nonuniform
  state,'' {\em Zh. Eksp. Teor. Fiz} {\bf 68} (1975) 175--185.

\bibitem{SwiftHTwo}
P.~C. Hohenberg and J.~B. Swift, ``Metastability in fluctuation-driven
  first-order transitions: Nucleation of lamellar phases,'' {\em Phys. Rev.}
  {\bf E52} (1995) 1828.

\bibitem{SwiftH}
J.~B. Swift and P.~C. Hohenberg, ``Hydrodynamic fluctuations at the convective
  instability,'' {\em Phys. Rev.} {\bf A15} (1977) 319.

\bibitem{Bates}
F.~S. Bates, J.~H. Rosedale, G.~H. Frederickson, and C.~J. Glinka,
  ``Fluctuation-induced first-order transition of an isotropic system to a
  periodic state,'' {\em Phys. Rev. Lett.} {\bf 61} (1988) 2229--2232.

\bibitem{Landau}
L.~D. Landau and E.~M. Lifshitz, {\em Statistical Physics}.
\newblock Pergamon Press, Oxford, 1969.

\bibitem{AmbjornOne}
J.~Ambjorn, Y.~M. Makeenko, J.~Nishimura, and R.~J. Szabo, ``Nonperturbative
  dynamics of noncommutative gauge theory,''
  \href{http://xxx.lanl.gov/abs/hep-th/0002158}{{\tt hep-th/0002158}}.

\bibitem{HarveyKraus}
J.~A. Harvey and P.~Kraus, ``D-branes as unstable lumps in bosonic open string
  field theory,'' {\em JHEP} {\bf 04} (2000) 012,
  \href{http://xxx.lanl.gov/abs/hep-th/0002117}{{\tt hep-th/0002117}}.

\bibitem{Yi:1999hd}
P.~Yi, ``Membranes from five-branes and fundamental strings from Dp branes,''
  {\em Nucl. Phys.} {\bf B550} (1999) 214,
  \href{http://xxx.lanl.gov/abs/hep-th/9901159}{{\tt hep-th/9901159}}.

\bibitem{Sen:1999md}
A.~Sen, ``Supersymmetric world-volume action for non-BPS D-branes,'' {\em JHEP}
  {\bf 10} (1999) 008, \href{http://xxx.lanl.gov/abs/hep-th/9909062}{{\tt
  hep-th/9909062}}.

\bibitem{Mukhi}
K.~Dasgupta, S.~Mukhi, and G.~Rajesh, ``Noncommutative tachyons,'' {\em JHEP}
  {\bf 06} (2000) 022, \href{http://xxx.lanl.gov/abs/hep-th/0005006}{{\tt
  hep-th/0005006}}.

\bibitem{Read98}
N.~Read, ``Lowest Landau level theory of the quantum Hall effect: The Fermi
  liquid - like state,'' {\em Phys. Rev.} {\bf B58} (1998) 16262,
  \href{http://xxx.lanl.gov/abs/cond-mat/9804294}{{\tt cond-mat/9804294}}.

\bibitem{Koulakov}
A.~A. Koulakov, M.~M. Fogler, and B.~I. Shklovskii, ``Charge density wave in
  two-dimensional electron liquid in weak magnetic field,'' {\em Phys. Rev.
  Lett.} {\bf 76} (1996) 499,
  \href{http://xxx.lanl.gov/abs/cond-mat/9508017}{{\tt cond-mat/9508017}}.

\bibitem{Moessner}
R.~Moessner and J.~T. Chalker, ``Exact results for interacting electrons in
  high Landau levels,'' {\em Phys. Rev.} {\bf B54} (1996) 5006,
  \href{http://xxx.lanl.gov/abs/cond-mat/9606177}{{\tt cond-mat/9606177}}.

\bibitem{Zirnbauer}
M.~R. Zirnbauer, ``Conformal field theory of the integer quantum Hall plateau
  transition,'' \href{http://xxx.lanl.gov/abs/hep-th/9905054}{{\tt
  hep-th/9905054}}.

\end{thebibliography}\endgroup
